\documentclass{jfm}

\usepackage{graphicx}

\usepackage{newtxtext}
\usepackage{newtxmath}
\usepackage{natbib}
\usepackage{hyperref}
\hypersetup{
    colorlinks = true,
    urlcolor   = blue,
    citecolor  = blue,
}
\usepackage{cleveref}

\newcommand{\RomanNumeralCaps}[1]
\linenumbers
\usepackage[justification=raggedright,singlelinecheck=false]{caption}

\usepackage[]{todonotes}

\title{Exact Coherent Structures of Sheared Double-Diffusive Convection}

\author{Van Duc Nguyen\aff{1}
  \and Chang Liu\aff{1}\corresp{\email{chang\_liu@uconn.edu}}}

\affiliation{\aff{1}School of Mechanical, Aerospace, and Manufacturing Engineering, University of Connecticut, Storrs, CT 06269, USA}

\begin{document}
\maketitle
\allowdisplaybreaks

\begin{abstract}
The interaction between shear and double-diffusive convection (DDC) in the diffusive regime (cold fresh water on top of hot salty water) plays an important role in the heat and mass transport of polar region oceans. This study computes exact coherent structures (ECS) of diffusive-regime DDC with a uniform background shear in a vertically wall-bounded flow layer. We focus on the shear-influenced regime and present two-dimensional (2D) ECS consisting of steady-state solutions and periodic orbits. The steady-state solutions include tilted convective rolls with various horizontal wavenumbers, and they are invariant under horizontal translation. All tilted convective roll states undergo saddle-node bifurcation, leading to a stable upper branch and an unstable lower branch, suggesting that they originate from the subcritical bifurcation of conduction base states. Hopf bifurcations appear on the stable upper branch of tilted convective rolls, leading to periodic orbits. Bifurcation diagrams for dimensionless parameters, including the Rayleigh number, the Prandtl number, and the diffusivity ratio, are established, suggesting subcritical behavior. Increasing shear strength stabilizes the 2D tilted convective roll, while these tilted convective rolls continue to exist in the limit of zero density ratio corresponding to sheared Rayleigh-B\'enard convection. Extension to the three-dimensional (3D) domain leads to 3D streamwise-elongated roll solutions; one of them originates from a subcritical bifurcation of the corresponding 2D roll solution. Chaotic solutions from direct numerical simulations generally visit neighborhoods of these steady or periodic solutions, and these visits leave an imprint on the flow statistics. 

\end{abstract}



\section{Introduction}
\label{sec:intro}
A fluid subjected to a gradient due to the combination of two scalars with different diffusivities, such as temperature and salinity, is often called double diffusive convection (DDC) \citep{radko2013double}. The DDC process plays a crucial role in a wide range of applications, from geophysics \citep{huppert1984double} and marine engineering \citep{bouffard2019convection,meiburg2024fluid} to astrophysics \citep{leconte2017condensation,garaud2018double,PhysRevFluids.6.030501}. There are two important regimes of DDC, including the diffusive regime, where cold and fresh water overlies warm and salty water, and the fingering regime in the reverse setting with warm and salty water on top of cold and fresh water. Both regimes of DDC lead to instabilities driven by the diffusivity difference between salt and temperature \citep{stern1960salt,radko2013double}, even when the density is overall stably stratified. This process plays an important role in the vertical mixing of temperature and salinity around the world's oceans \citep{schmitt1994double}.

This work focuses on the diffusive regime of DDC (or diffusive convection), which is relevant to the ocean in polar regions. DDC in the diffusive regime offers important insights into large-scale thermohaline circulations and the climate of polar regions \citep{aagaard1989synthesis,bebieva2017relationship,bebieva2019regulation}. For example, \citet{timmermans2020understanding} reviewed that diffusive staircases act as a critical process in the upward transfer of heat in both the warm Atlantic waters and the rapidly changing Arctic regions. Diffusive convection not only enhances sea ice formation in the winter months \citep{bebieva2019regulation}, but also contributes to the summer warming of subsurface cold Antarctic winter water \citep{giddy2023vertical}. Exploring and understanding the dynamic properties of DDC in these areas is crucial for modeling global climate change \citep{ferrari2009ocean,rosevear2025does}.

Shear flow is also observed accompanying diffusive convection, as shown in field measurements of temperature, salinity, and velocity profiles within Antarctic ice shelf cavities using underwater robotics \citep{schmidt2023heterogeneous,davis2023suppressed,washam2023direct} or borehole drilling \citep{minowa2021thermohaline}.  These field measurements not only show the high temperature and high salinity at the bottom corresponding to diffusive convection, but also indicate that the flow speed differs over vertical position (i.e., shear flows) within the ice shelf cavities. Shear flows and DDC also play an important role in influencing the ice melting rate \citep{toppaladoddi2021nonlinear,ravichandran2022combined,du2024physics,rosevear2025does}. For example, shear flows from the lateral turbulent mixing allow warm water to reach sloped ice surfaces and promote melting \citep{schmidt2023heterogeneous}, while DDC controls ice shelf basal melting and turbulent mixing at low current speeds \citep{rosevear2021role}. Moreover, \citet{rosevear2022regimes} constructed a regime diagram to demonstrate a transition between double-diffusive, stratified, and shear-driven melting regimes characterized by the background shear flow strength and temperature beneath ice sheets. These observations motivate us to analyze how shear flows interact with the diffusive-regime DDC.

The interaction between the shear and diffusive regime DDC can lead to new instabilities and staircase formation. For example, \citet{radko2016thermohaline} demonstrated that diffusive DDC and a sinusoidal background velocity (i.e., Kolmogorov flow) can lead to instabilities up to the density ratio $\Lambda\approx10$ and Richardson number $Ri\approx10^3$. Here, the density ratio $\Lambda$ represents the ratio of density variations induced by background salinity and temperature gradients, and the Richardson number $Ri$ measures the relative relationship between the buoyancy force and the shear force in the fluid flow. This instability, called thermohaline-shear instability \citep{radko2016thermohaline}, appears in parameter regimes outside of the narrow parameter regime of diffusive instability $1<\Lambda<(Pr+1)/(Pr+\tau)$ \citep{walin1964note} or $1<\Lambda<1.14$ with typical oceanic parameters $(Pr,\tau)=(7,0.01)$. Here, the Prandtl number $Pr$ represents the ratio of fluid viscosity to thermal diffusivity, and $\tau$ denotes the diffusivity ratio between salinity and temperature. This thermohaline-shear instability also appears at the Richardson number ($Ri$) much higher than the threshold $Ri=1/4$, where $Ri<1/4$ is a necessary condition for dynamical instability in single-component non-dissipative stratified shear flows based on the Miles-Howard theorem \citep{miles1961stability,howard1961note}. Direct numerical simulations (DNS) also show that this thermohaline-shear instability will evolve into thermohaline staircases \citep{radko2016thermohaline}, and thus this instability is hypothesized to be responsible for thermohaline staircase formation around the world ocean in diffusively and dynamically stable regimes. Such thermohaline-shear instability also persists when the background shear flow has a periodic time dependence \citep{radko2019thermohaline,brown2019initiation,kochnev2025stability}. 

Within a vertically bounded domain, diffusive convection interacting with a linear background velocity can still lead to staircase formation \citep{yang2022layering}, although linear instabilities were not identified. The bounded domain may be more suitable for modeling seawater within an ice-shelf cavity, where the ice-water interface will confine the length scale in certain directions \citep{gayen2016simulation,mondal2019ablation,rosevear2021role,guo2025effects}. Moreover, such a uniform shear flow configuration resembles the canonical plane Couette flow (PCF) \citep{gibson2008visualizing}, which is linearly stable for any Reynolds number \citep{romanov1973stability} due to the disappearance of inflection points in the background velocity profile. Within the diffusive instability regime, the Couette flow significantly enhances the transient growth of diffusive convection and stabilizes the effects of diffusive instability \citep{lu2024convective, lu2025transient}. Outside of the diffusive instability regime, \citet{yang2022layering} investigated diffusive-regime DDC in a vertically bounded domain with a uniform background shear, referred to as sheared double-diffusive convection (SDDC). Similar to PCF, the nonlinear and turbulent behavior of SDDC needs to be triggered by a finite amplitude disturbance \citep{yang2022layering}. DNS results show flow behaviors of multiple layers and layer merging at Richardson number $Ri=1$ and density ratio $\Lambda=2$ \citep{yang2022layering}. Furthermore, \citet{li2024double} provided a more comprehensive understanding of vertical transport in the SDDC over a wide range of Richardson numbers $Ri\in[0.3,4]$. This study determined the final states as the laminar state, one convective layer, or layering states \citep{li2024double} depending on the Richardson number. The laminar state is dominant under the weak shear associated with high $Ri$, while layering structures are highly sensitive to the amplitude and vertical wavenumber of initial disturbances as well as the running time of simulations \citep{li2024double}. Despite the long lifetime of multi-layer states, they will usually transition to the one convective layer state after sufficiently long simulations via the layer merging mechanism \citep{chong2020cafe,li2024double}.

To improve our understanding of SDDC, it is crucial to determine the nonlinear solutions and their stability in SDDC. These nonlinear solutions are usually referred to as exact coherent structures (ECS) \citep{kawahara2012significance,graham2021exact}, which can complement DNS-based statistical analysis \citep{chong2020subcritical,yang2022layering,li2024double} to advance our understanding of the skeleton of turbulent or chaotic behavior and the transition to turbulence. ECS encapsulates many characteristics of coherent structures in turbulent flows and provides insights into the mechanisms of self-sustaining turbulence \citep{nagata1990three,gibson2009equilibrium,waleffe2009exact,schneider2010localized,langham2020stably,beneitez2024transition}. Their dynamic influence on flows through invariant manifolds is observable through experiments and can be examined in detail using DNS \citep{barkley2005computational,reetz2019exact,blass2020flow}. The utility of ECS has been demonstrated in exploring the pathway of transition to turbulence and its underlying characteristics. In particular, SDDC associated with a linear background velocity represents the interaction of two subcritical systems (i.e., PCF and diffusive convection), whose ECS remains unexplored. 

Exact coherent structures have been analyzed in several related flow configurations, such as PCF, stratified PCF, Rayleigh-B\'enard convection (RBC), sheared RBC, and DDC. After the first ECS for PCF was identified by \citet{nagata1990three}, an extended series of ECS for PCF was then computed by \citet{gibson2008visualizing,gibson2009equilibrium,halcrow2008charting}. The periodic orbits of PCF were shown to reproduce the turbulent mean velocity and root mean square of velocity fluctuations \citep{kawahara2001periodic}, and 3D multiscale ECS of RBC also agrees well with mean temperature and root-mean-squares of fluctuations in turbulent states \citep{motoki2021multi}. The lower branch ECS of PCF in the form of streamwise streaks and rolls was shown to have a single unstable eigenvalue at all Reynolds numbers, which suggests that this state controls the transition to turbulence of PCF \citep{wang2007lower}. For stratified PCF, \citet{olvera2017exact} and \citet{langham2020stably} continued an equilibrium solution EQ7 (a two-layer equilibrium solution from the ECS list of PCF \citep{gibson2009equilibrium}) from unstratified to stratified PCF. The influence of a uniform shear on the wavy instability of longitudinal convection rolls of Rayleigh-B\'enard convection was also investigated by ECS \citep{clever1992three}. Moreover, convection within an inclined layer can generate a background shear flow, which interacts with convection rolls to generate subharmonic standing waves, wavy rolls, knots, transverse oscillations \citep{reetz2020invariant,reetz2020invariant_part2}, and spatially localized structures \citep{wen2019moderate,li2025traveling}. Studies by \citet{zheng2024natural,zheng2024natural2,zheng2025natural} on natural convection in a vertical channel found equilibrium solutions in the form of two- and three-dimensional counter-rotating convection rolls, which transition to chaos through a sequence of bifurcations of periodic orbits. Moreover, the spontaneous generation of shear flow can result from a secondary bifurcation of convection rolls or elevator mode \citep{howard1986large,rucklidge1996analysis,liu2022staircase,liu2024fixed}, and this shear flow can tilt the convective flow structure and reduce heat or mass transport. Bifurcation analysis of the diffusive regime DDC also shows the generation of standing waves and traveling waves from the Hopf bifurcation with $O(2)$ symmetry \citep{knobloch1986oscillatory,tuckerman2001thermosolutal,liu2022single}.

Our goal in this study is to determine the ECS for SDDC in a wall-bounded flow layer. We consider a linear background shear flow that is closely connected to the PCF, a canonical flow that has the subcritical transition \citep{gibson2009equilibrium}. The no-slip boundary condition is used for the velocity fluctuations around the background shear flow, and this boundary condition is relevant to lab-scale experiments \citep{hage2010high} and ice shelf basal melting \citep{guo2025effects}. We will start by presenting a series of DNS to identify the evolution of one-layer convection in SDDC and classify the parameter regime in this work as the shear-influenced regime \citep{li2024double}. We then found the two-dimensional ECS of SDDC, including equilibrium solutions and periodic orbits. Equilibrium solutions show the subcritical coexistence of one-layer tilted roll states with various horizontal wavenumbers. We built bifurcation diagrams over the Rayleigh number, the Prandtl number, and the diffusivity ratio, which exhibit subcritical behavior. Here, lower branch solutions are unstable, while upper branch solutions are stable near saddle-node bifurcation points. We also found mixed modes connecting the lower and upper branches of equilibrium solutions. The upper branch solutions lose stability due to a Hopf bifurcation as the Rayleigh number increases, leading to periodic orbits. Increasing the shear strength will stabilize the tilted convective rolls, while some tilted roll solutions continue to exist in the zero density ratio limit corresponding to sheared Rayleigh-B\'enard convection. Three-dimensional (3D) roll states exist, with one of them bifurcating from the corresponding two-dimensional (2D) state through a subcritical bifurcation when extending the spanwise domain size.

The rest of this paper is organized as follows. We present the governing equations and numerical methods for SDDC in section \ref{sec:method}. In section \ref{sec:dns}, we conduct direct numerical simulations of long time to motivate finding ECS in the form of one-layer convection. We then present two-dimensional ECS, including the equilibrium solutions in Section \ref{sec:equilibrium} and periodic orbits in Section \ref{sec:periodic_orbit}. The effects of dimensionless parameters are analyzed in Section \ref{sec:parameter_continuation}, and the 3D convective roll states are discussed in Section \ref{sec:3D_effect}. We conclude this paper with a discussion of future directions in Section \ref{sec:conclusion}.

\section{Problem formulation}\label{sec:method}
\begin{figure}
    \centering
    \includegraphics[width=0.6\linewidth]{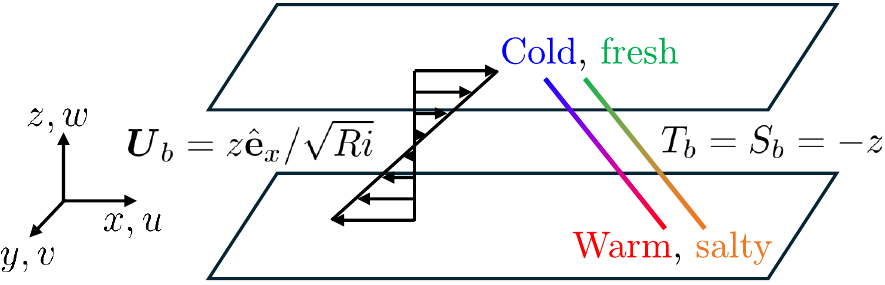}
    \caption{Illustration of the flow configuration with the computational domain as $L_x=2\pi$, $L_z=1$ (with vertical coordinates $z\in [-0.5,0.5]$), and varying spanwise size $L_y$. Laminar base state velocity is $\boldsymbol{U}_b=z\hat{\mathbf{e}}_x / \sqrt{Ri}$, where $Ri$ represents the Richardson number. The conduction base state temperature and salinity profiles are $T_b=S_b=-z$. }
    \label{fig:domain}
\end{figure}

\subsection{Governing equations}

We consider an incompressible viscous fluid flow layer between two parallel plates with a height $H$. Two plates move in the $x$ streamwise direction with a constant velocity $z^* \hat{\mathbf{e}}_x$, leading to a linear background velocity profile $\boldsymbol{U}_b^*=z^* \hat{\mathbf{e}}_x$, where $z^*$ is the coordinate in the $z$ vertical direction and $\hat{\mathbf{e}}_x$ denotes the unit vector in the $x$ streamwise direction. The superscript `$*$' indicates the dimensional quantities throughout this paper. The shear strength is represented by the wall velocity difference $U_w^*=|z_{max}^*-z_{min}^*|$. Fluid density is influenced by temperature and salinity using the equation of state $\rho^*(T^*, S^*)=\rho^*_{0}\left[1-\beta_T\left(T^*-T_{0}^*\right)+\beta_S\left(S^*-S_{0}^*\right)\right]$ linearized around $\rho^*_0(T^*_0,S^*_0)$, where the subscript `0' represents the reference values for linearization. Here, $\beta_T$ and $\beta_S$ are, respectively, the expansion coefficient of temperature and the contraction coefficient of salinity. Under the Oberbeck–Boussinesq (OB) approximation, the dimensional governing equations for DDC \citep{li2023wall} are expressed by
\begin{subequations}
\label{eq:NS_dimensional_all}
\begin{align}
\nabla^* \cdot \boldsymbol{U}^{*}&=0,\\
\frac{\partial \boldsymbol{U}^{*}}{\partial t^*}+\boldsymbol{U}^{*} \cdot \nabla^* \boldsymbol{U}^{*}&=-\nabla^* P^{*}+v \nabla^{*2} \boldsymbol{U}^{*}+g\left[\beta_{T} (T^{*}-T^*_0)-\beta_{S} (S^{*}-S^*_0)\right] \hat{\mathbf{e}}_{z},\label{eq:NS_momentum} \\
\frac{\partial T^{*}}{\partial t^*}+\boldsymbol{U}^{*} \cdot \nabla^* T^{*}&=\kappa_{T} \nabla^{*2} T^{*}, \\
\frac{\partial S^{*}}{\partial t^*}+\boldsymbol{U}^{*} \cdot \nabla^* S^{*}&=\kappa_{S} \nabla^{*2} S^{*},
\end{align}
\end{subequations}
where $\boldsymbol{U}^*$, $P^*$, and $t^*$ are the dimensional fluid velocity, pressure, and time, respectively. In \eqref{eq:NS_dimensional_all}, $\nu$, $\kappa_T$, and $\kappa_S$ denote the kinematic viscosity of the fluid, the thermal diffusivity, and the salinity diffusivity, respectively. In \eqref{eq:NS_momentum}, $g$ represents the gravitational acceleration, and $\hat{\mathbf{e}}_z$ denotes the unit vector in the $z$ vertical direction.

We non-dimensionalize the length quantity as $z=z^*/H$ with $H$ as the layer height, which leads to a non-dimensional gradient operator represented as $\nabla=\nabla^* H$. Temperature and salinity are non-dimensionalized as $T=(T^*-T^*_0)/\Delta T^*$ and $S=(S^*-S^*_0)/\Delta S^*$, where $\Delta T^*=|T_{bot}^*-T_{top}^*|$ and $\Delta S^*=|S_{bot}^*-S_{top}^*|$ are, respectively, the temperature and salinity differences between the bottom (`bot' subscript) and top (`top' subscript) planes. Velocity is non-dimensionalized by the free-fall velocity associated with the temperature property; i.e., $U=U^*/u_{\text{ff}}$ with $u_{\text{ff}}:=\sqrt{g\beta_T \Delta T^* H}$. Correspondingly, time is non-dimensionalized by $t=t^*u_{\text{ff}}/H$. Then, we obtain the non-dimensional Oberbeck-Boussinesq equations for the sheared double-diffusive convection (SDDC) as:
\begin{subequations}
\label{eq.sddc_nondimensional}
\begin{align}
\nabla\cdot \boldsymbol{U} &= 0,\\
\frac{\partial\boldsymbol{U}}{\partial t} +\boldsymbol{U}\cdot\nabla\boldsymbol{U} &= -\nabla P + \sqrt{\frac{Pr}{Ra}}\nabla^2\boldsymbol{U} + \left(T-\Lambda S\right)\hat{\mathbf{e}}_z,\label{eq.sddc_nondimensional.momentum}\\
\frac{\partial T}{\partial t} +\boldsymbol{U}\cdot\nabla T &= \frac{1}{\sqrt{Pr Ra}}\nabla^2 T,\\
\frac{\partial S}{\partial t} +\boldsymbol{U}\cdot\nabla S &= \frac{\tau}{\sqrt{Pr Ra}}\nabla^2 S.
\end{align}
\end{subequations}
In \eqref{eq.sddc_nondimensional}, we have dimensionless parameters, including the Rayleigh number ($Ra$), the Prandtl number ($Pr$), the diffusivity ratio ($\tau$), the density ratio ($\Lambda$), and the Richardson number ($Ri$):
\begin{equation}
Ra=\frac{g\beta_T \Delta
 T^* H^3}{\kappa_T\nu}, \quad Pr=\frac{\nu}{\kappa_T}, \quad \tau=\frac{\kappa_S}{\kappa_T}, \quad \Lambda=\frac{\beta_S\Delta S^*}{\beta_T\Delta T^*}, \quad Ri=\frac{g\beta_T \Delta
 T^* H}{U_w^{*2}}.
\end{equation}
The density can be normalized by the density difference between the bottom and top walls $\Delta \rho^*=\rho_{bot}^*-\rho_{top}^*=\rho^*_0(\beta_S\Delta S^*-\beta_T \Delta T^*)$ \citep{yang2022layering}, leading to a non-dimensional density as follows: 
\begin{equation}
    \rho:=\frac{\rho^*-\rho^*_0}{\Delta \rho^*} =\frac{\rho^*_0[\beta_S (S^*-S^*_0)-\beta_T (T^*-T^*_0)]}{\rho^*_0(\beta_S\Delta S^*-\beta_T\Delta T^*)}=\frac{\Lambda S-T}{\Lambda-1}. 
\end{equation}

In this study, we investigate SDDC in a flow layer between two parallel plates with temperature $T=\mp 0.5$ and salinity $S=\mp 0.5$ at $z=\pm 0.5$, corresponding to the diffusive regime DDC, leading to $\rho=\mp 0.5$ at $z=\pm 0.5$, the top and bottom walls. The top and bottom walls $z=\pm 0.5$ are moving with velocities $\pm0.5\hat{\mathbf{e}}_x / \sqrt{Ri}$ (see figure \ref{fig:domain}), where the Richardson number is related to the shear strength $Ri^{-1/2}=U_w$ with $U_w$ as the non-dimensional wall velocity difference. 

We then decompose velocity, temperature, salinity, and pressure into two components: a steady base state ($\boldsymbol{U}_b, T_b, S_b, P_b$) and fluctuations around this base state ($\boldsymbol{u}, \theta, s, p$):
\begin{equation}
    \boldsymbol{U}=\boldsymbol{U}_b+\boldsymbol{u}, \quad T=T_b+\theta, \quad S=S_b+s, \quad P=P_b+p.
    \label{eq:base_state}
\end{equation}
This strategy is effective in implementing our code, which will be discussed in subsection \ref{sub:channelflow-ddc}. The base state consists of a laminar base flow ($\boldsymbol{U}_b$), linear scalar profiles ($T_b$, $S_b$), and a mean pressure ($P_b$): 
\begin{equation}
\boldsymbol{U}_b=z\hat{\mathbf{e}}_x / \sqrt{Ri}, \quad T_b=S_b=-z, \quad P_b=(\Lambda-1) z^2/2.
\label{eq.base_state}
\end{equation}
The equation \eqref{eq.sddc_nondimensional} can be rewritten in terms of fluctuations as follows: 
\begin{subequations}
\label{eq.sddc_fluctuation}
\begin{align}
\nabla\cdot \boldsymbol{u} &= 0,\\
\frac{\partial\boldsymbol{u}}{\partial t} + U_b \frac{\partial \boldsymbol{u}}{\partial x} + w\frac{d U_b}{d z}\hat{\mathbf{e}}_x +\boldsymbol{u}\cdot\nabla\boldsymbol{u} &= -\nabla p + \sqrt{\frac{Pr}{Ra}}\nabla^2\boldsymbol{u} + \left(\theta-\Lambda s\right)\hat{\mathbf{e}}_z,\\
\frac{\partial \theta}{\partial t} + U_b \frac{\partial\theta}{\partial x} +\boldsymbol{u}\cdot\nabla \theta - w &= \frac{1}{\sqrt{Pr Ra}}\nabla^2 \theta,\\
\frac{\partial s}{\partial t} + U_b \frac{\partial s}{\partial x} +\boldsymbol{u}\cdot\nabla s - w &= \frac{\tau}{\sqrt{Pr Ra}}\nabla^2 s.
\end{align}
\end{subequations}
Here, $U_b=z/\sqrt{Ri}$ and  $\boldsymbol{u}=[u,v,w]^\text{T}$ with $u$, $v$, and $w$ representing the horizontal ($x$ and $y$ axes) and vertical ($z$ axis) velocity fluctuation components, as shown in Figure \ref{fig:domain}.

In this study, we use the no-slip boundary condition for velocity fluctuations at the top and bottom plates, while temperature and salinity are fixed at the walls:
\begin{equation}
    \boldsymbol{u}(x,y,z=\pm0.5,t)=\boldsymbol{0},\quad
    \theta(x,y,z=\pm0.5,t)=s(x,y,z=\pm0.5,t)=0.
\end{equation}
A periodic boundary condition is applied for all variables in the horizontal $x$ and $y$ directions.

\subsection{Numerical methods}\label{sub:channelflow-ddc}

We use a modified parallelized-spectral code of \textit{Channelflow 2.0} \citep{channelflow2} to conduct direct numerical simulations and compute ECS based on the governing equations \eqref{eq.sddc_fluctuation}. To simplify, we call this modified \textit{Channelflow} code ``ChFlow-DDC" throughout this paper. The ChFlow-DDC employs the Fourier-Fourier-Chebyshev spectral method in $x$, $y$, and $z$ directions for representing variable fields in the spatial domain \citep{gibson2008visualizing,reetz2020invariant}. For 2D results presented in \cref{sec:dns,sec:equilibrium,sec:periodic_orbit,sec:parameter_continuation}, we fix a thin spanwise layer $L_y=5\times 10^{-3}$ with a corresponding grid $N_y=10$ to suppress the instability in the spanwise variation of the velocity field, implying a flow configuration without three-dimensional effects. This approach also allows us to continue the two-dimensional (2D) solutions to the three-dimensional (3D) solutions by increasing the spanwise domain size $L_y$, and the 3D effect will be discussed in section \ref{sec:3D_effect}. We fix the streamwise and vertical lengths $L_x\times L_z=2\pi\times 1$ that are discretized by collocation points $N_x\times N_z=128\times97$ unless otherwise mentioned. A detailed validation for ChFlow-DDC against published results \citep{yang2015salinity,langham2020stably,zheng2024natural} is presented in Appendix \ref{app:validations}. We also compare our results within a thin spanwise domain $L_y$ with the 2D simulation results obtained by Dedalus \citep{burns2020dedalus}, which shows negligible difference; see Appendix \ref{app:mesh_ind_study}.

To construct the ECS underlying the complex spatio-temporal dynamics of flow, the shooting-based Newton-Raphson method coupled with the matrix-free Krylov method is applied in ChFlow-DDC to determine the state vector $\boldsymbol{\zeta}(t)=\begin{bmatrix}\boldsymbol{u}\\ \theta \\s\end{bmatrix}$ that satisfies 
\begin{equation}
    \mathcal{G}(\boldsymbol{\zeta}) = \sigma \mathcal{F}^T(\boldsymbol{\zeta}) - \boldsymbol{\zeta} = 0,
    \label{eq.invariant_solution}
\end{equation} 
where $\sigma$ is the symmetry operator and $\mathcal{F}^T$ is the time-$T$ forward integration of the governing equations in \eqref{eq.sddc_fluctuation}. The sequential Krylov vectors are constructed via time-stepping of initial disturbances \citep{kelley2003solving}, which are then employed in a hook-step trust-region optimization that significantly enhances the convergence radius \citep{viswanath2007recurrent}. An ECS is deemed to be a converged ECS when the residual norm of (\ref{eq.invariant_solution}) falls below $10^{-13}$. These converged solutions are then extended parametrically as a continuation across dimensionless parameters to generate bifurcation diagrams. A mesh independence study is performed (see Appendix \ref{app:mesh_ind_study}) to demonstrate the mesh convergence of our current mesh for both DNS and equilibrium solutions.

\subsection{Symmetries of system}
The organization of flow patterns and their internal transitions is governed by symmetry and symmetry-breaking bifurcation \citep{crawford1991symmetry}. To identify the flow configuration of ECS in space, we employ a symmetry system based on $z$-reflection (\ref{eq:sym_reflection_z}), $x$-reflection (\ref{eq:sym_reflection_x}), and a symmetry family of translation in the streamwise (\ref{eq:sym_translation_x}) and spanwise directions (\ref{eq:sym_translation_y}):
\begin{subequations}
\label{eq:sym_group}
\begin{align}
\pi_{z}[u, v, w, \theta, s](x, y, z) &\equiv [u, v,-w, -\theta, -s](x, y,-z), \label{eq:sym_reflection_z}\\
\pi_{x}[u, v, w, \theta, s](x, y, z) &\equiv [-u, v, w, \theta, s](-x, y, z), \label{eq:sym_reflection_x}\\
\tau_x(\lambda_x)[u, v, w, \theta, s](x, y, z) &\equiv [u, v, w, \theta, s](x+\lambda_x, y, z), \label{eq:sym_translation_x}\\
\tau_y(\lambda_y)[u, v, w, \theta, s](x, y, z) &\equiv [u, v, w, \theta, s](x, y+\lambda_y, z).\label{eq:sym_translation_y}
\end{align}
\end{subequations}
Here, $z$-reflection symmetry $\pi_z$ in (\ref{eq:sym_reflection_z}) corresponds to vertical reflection via $z\rightarrow -z$, which is associated with the change of the sign of scalar fluctuation fields in convection problems \citep{langham2020stably,zheng2024natural}: $\theta\rightarrow-\theta$ and $s\rightarrow-s$. The $x$-reflection symmetry $\pi_x$ in (\ref{eq:sym_reflection_x}) \citep{liu2022staircase} reflects the flow configuration, $x\rightarrow-x$, through the plane $x=0$. The $x$-translation symmetry $\tau_x(\lambda_x)$ in (\ref{eq:sym_translation_x}) and $y$-translation symmetry $\tau_y(\lambda_y)$ in (\ref{eq:sym_translation_y}) are considered due to the horizontally periodic boundary conditions. The shifting intervals ($\lambda_x$,$\lambda_y$) are arbitrary, such that $0\leq\lambda_{x}\leq L_{x}$ and $0\leq \lambda_y\leq L_y$. In a context without a specific value of $\lambda_x$ or $\lambda_y$, symmetries $\tau_x(L_x)$ or $\tau_y(L_y)$ imply trivial symmetries for every solution of any system with the $x$- and $y$-periodic directions.  A collection of symmetries represented by $\langle \pi_z,\pi_x,\tau_x(\lambda_x),\tau_y(\lambda_y) \rangle$ will be applied together as a constraint to identify solutions. The symmetry $\tau_y(\lambda_y)$ will be ignored for 2D solutions.

\section{Direct numerical simulation}\label{sec:dns}
\subsection{Evolution to one-layer convection}\label{sec:dns.1}
\begin{figure}
\centering
\includegraphics[width=1\linewidth]{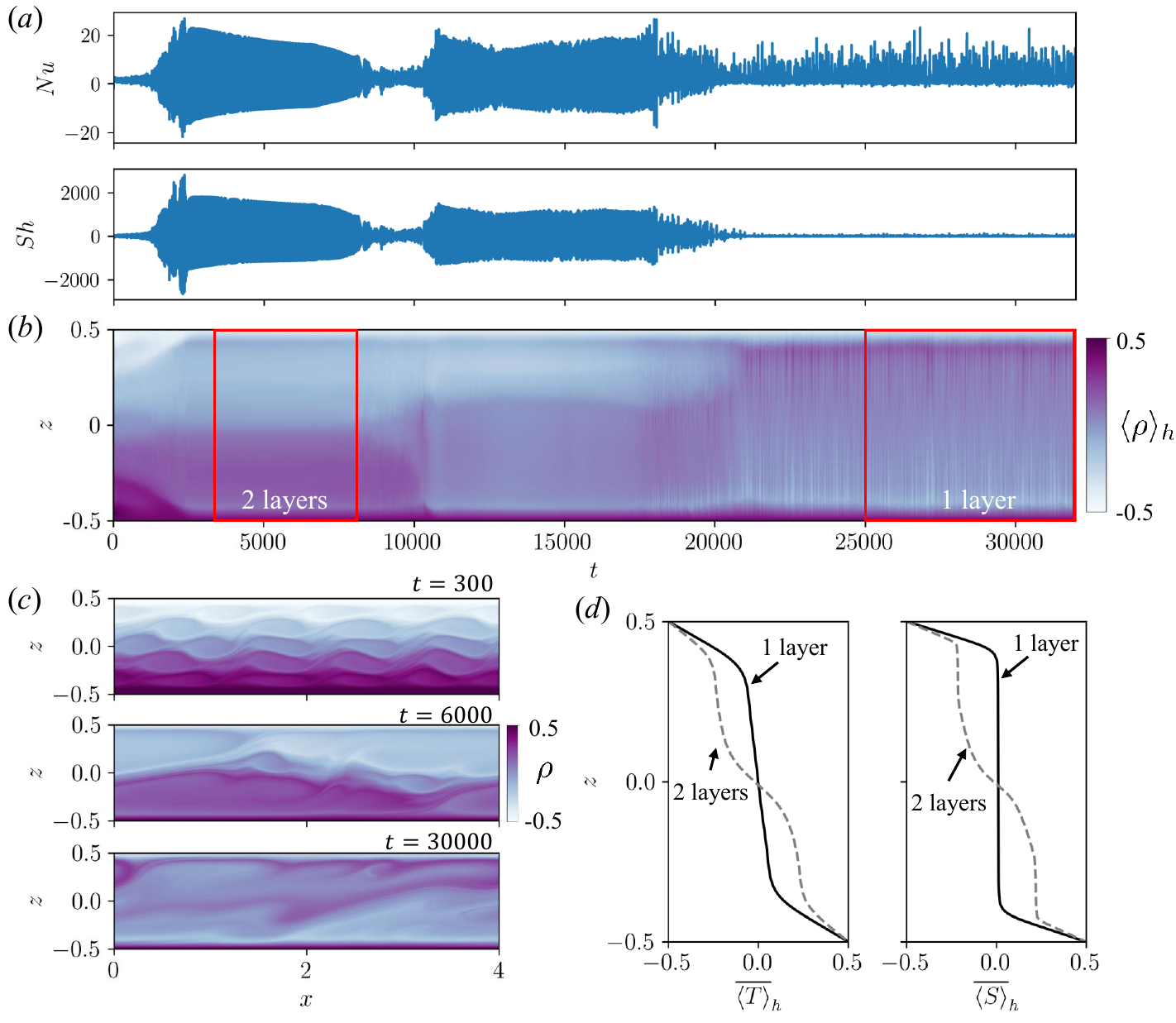}
\caption{Direct numerical simulation of SDDC in a vertically wall-bounded flow layer with parameters $(Ra,Ri,Pr,\tau,\Lambda)=(10^6,1,10,0.01,2)$. (a) Time evolution of $Nu$ and $Sh$. (b) Time history of horizontally-averaged total density $\langle \rho\rangle_h$ is plotted. (c) Distributions of the instantaneous total density field ($\rho$) at three different times illustrate initial development ($t=300$) and density stratification with two layers ($t=6000$) and one layer ($t=30000$) as the final state. (d) Time and horizontally averaged scalar profiles of total temperature $\overline{\langle T\rangle}_{h}$ and salinity $\overline{\langle S\rangle}_{h}$. Dashed gray lines correspond to mean profiles averaged over $t\in [3000,8000]$ displaying two layers, while solid black lines are averaged over $t\in [25000,32000]$ associated with the one-layer state. These time-averaging domains correspond to two red boxes in panel (b). See supplementary movie 1.}	
\label{fig:dns_sddc}
\end{figure}

We start by presenting direct numerical simulation results to identify the final state and analyze the effect of no-slip compared with stress-free boundary conditions \citep{yang2022layering,li2024double}. Now we start with a 2-D DNS for SDDC with parameters $(Ra,Ri,Pr,\tau,\Lambda)=(10^6,1,10,0.01,2)$ corresponding to case 4 of \citet[Table 1]{yang2022layering} and case 3 of \citet[Table 1]{li2024double}. In this simulation, we use a grid resolution of $N_x\times N_z=768\times385$ for a domain size of $L_x\times L_z=4\times 1$ that is similar to the setup of \citet{li2024double}. All the initial disturbances for the velocity, temperature, and salinity fields were established consistently with \citet{yang2022layering} for a direct comparison, except for the velocity boundary condition on the walls. Here, we use the no-slip velocity boundary condition instead of the stress-free boundary condition employed by \citet{yang2022layering} and \citet{li2024double}.

Figure \ref{fig:dns_sddc} shows the density stratification and layer merging behaviors over time obtained from DNS. In particular, figure \ref{fig:dns_sddc}(a) shows the time history of the Nusselt number ($Nu$) and the Sherwood number ($Sh$), which are defined as follows:
\begin{subequations}
    \label{eq:Nu_Sh}
\begin{align}
    Nu=\left(\frac{\langle w^*T^*\rangle_h-\kappa_T\langle\partial_{z^*} T^*\rangle_h}{\kappa_T \Delta T^* H^{-1}} \right)_{z^*=0}=\left(\sqrt{Pr Ra}\langle wT\rangle_h - \langle \partial_z T\rangle_h \right)_{z=0},\\
    Sh=\left(\frac{\langle w^*S^*\rangle_h-\kappa_S\langle\partial_{z^*} S^*\rangle_h}{\kappa_S \Delta S^* H^{-1}} \right)_{z^*=0}=\left(\frac{\sqrt{Pr Ra}}{\tau}\langle wS\rangle_h - \langle \partial_z S\rangle_h \right)_{z=0}. 
\end{align}
\end{subequations}
The $Nu$ and $Sh$ measure the total flux due to convection and conduction, normalized by the corresponding conduction flux. Here, $Nu$ and $Sh$ are taken at the mid-plane $z=0$ that is common for convection problems \citep{yang2022layering,zheng2024natural,rosevear2025does}. At the beginning of DNS, multiple layers of convective rolls appear as an initial development; see the horizontally averaged total density $\langle \rho\rangle_h$ in figure \ref{fig:dns_sddc}(b) and snapshot $t=300$ in figure \ref{fig:dns_sddc}(c). Convective fluxes ($Nu$, $Sh$) gradually rise during this stage. This early behavior is also captured in DNS for SDDC by \citet{yang2022layering,li2024double}, which creates a mean density profile of multiple layers in the time range of $t\lessapprox 1000$ (not shown in figure \ref{fig:dns_sddc}). After that, the behavior of merging layers occurs (see figure \ref{fig:dns_sddc}(b)), leading to two layers and one layer represented by snapshots $t=6000$ and $t=30000$ in figure \ref{fig:dns_sddc}(c). Figure \ref{fig:dns_sddc}(d) then shows the horizontally and time-averaged profiles of total temperature and total salinity, where two layers and one layer are obtained from intervals $t\in[3000,8000]$, and $t\in[25000,32000]$, respectively, as indicated by the red boxes in figure \ref{fig:dns_sddc}(b). During the two-layer stage, $Sh$ has an extremely high amplitude larger than $\mathcal{O}(10^3)$, but they tend to decrease in the final one-layer state. This high-amplitude convective flux of salinity resembles the findings of \citet{yang2022layering} and \citet{ li2024double}. The transitional stage from two layers to one layer occurs over a long time. The one-layer stage has much smaller fluxes. Interestingly, both $Nu$ and $Sh$ exhibit chaotic fluctuations, but the two-layer stage always has a balance of negative and positive amplitudes.

Our findings indicate that the behaviors of layering and layer merging are robust in SDDC under the no-slip versus stress-free boundary conditions. The qualitative behaviors of layering and layer merging are similar to those observed by \citet{yang2022layering}. Our results display two layers with the no-slip velocity boundary condition, while three-layer behavior was observed in SDDC using the stress-free boundary condition \citep{yang2022layering}. We found that time-averaged convective fluxes $(\overline{Nu},\overline{Sh})=(3.05,5.62)$ with the no-slip boundary condition are smaller than $(\overline{Nu},\overline{Sh})=(6.19,18.5)$ \citep[Table 1]{li2024double} associated with the stress-free boundary condition. A comparison between no-slip and stress-free boundary conditions in salt-finger convection also shows similar conclusions. For example, \citet{yang2016vertically} demonstrated that the flow morphology of the two scalars within salt-finger convection is similar when using no-slip and stress-free velocity boundary conditions. In this case, convective fluxes ($Nu$, $Sh$) using the no-slip boundary condition are slightly smaller, but the exponents of the scaling laws are the same \citep{yang2016vertically}. Moreover, \citet{liu2022staircase} showed that steady-state solutions of salt-finger convection under stress-free boundary conditions have a higher $Sh$ than steady-state solutions under no-slip boundary conditions, although they show qualitatively similar flow structures. 

Moreover, the DNS results in figure \ref{fig:dns_sddc} indicate that one-layer convection is a common final state in DDC. Multiple-layer convection may be considered the transient stage due to computational limitations. This behavior was also captured by DNS in other studies \citep{chong2020cafe,li2024double}. One-layer convection is also a typical flow structure that is often observed in convection, along with findings of ECS \citep{gao2013transition,gao2015chaotic,reetz2019exact,wen2020steady,wen2022steady,motoki2021multi,liu2022single,zheng2024natural,zheng2024natural2,zheng2025natural}. Therefore, we will focus on constructing the ECS of the one-layer convection for SDDC in the present study. Actually, we attempted to search for ECS in the form of multiple layers in the vertical direction, but we have not found a converged solution resembling multiple layers. 

\subsection{Flow regime classification}\label{sec:dns.2}
\begin{figure}
\centering
\includegraphics[width=1\linewidth]{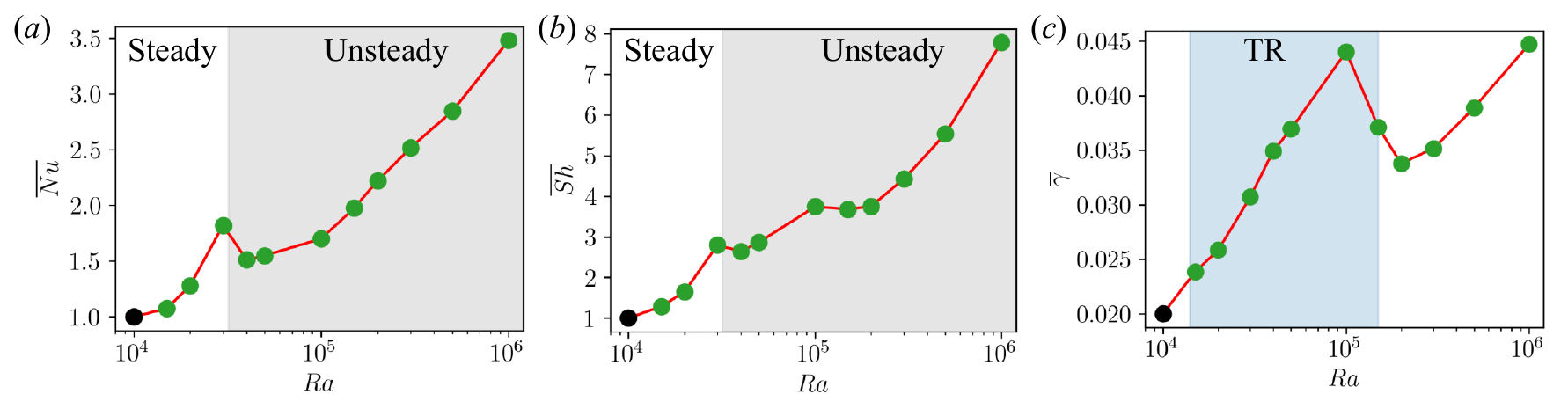}
\caption{Statistics of DNS data for SDDC. The Rayleigh number $Ra$ versus the time-averaged (a) Nusselt number $\overline{Nu}$, (b) Sherwood number $\overline{Sh}$, and (c) total flux ratio $\overline{\gamma}=\Lambda \tau \overline{Sh}/\overline{Nu}$. The black solid circles represent the laminar (trivial) flow regime, while the green solid circles represent the non-trivial flow regime. The white and gray zones in panels (a, b) indicate the steady and unsteady flow behavior, respectively. The blue zone in panel (c) indicates where tilted rolls (TR) are observed.}	
\label{fig:dns_flow_pattern}
\end{figure}

To find the interval of $Ra$ for the onset of non-trivial behavior of SDDC, we have conducted a series of DNS over a wide range of $Ra\in [10^4,10^6]$. In particular, figure \ref{fig:dns_flow_pattern} shows statistics from DNS for SDDC with dimensionless parameters fixed at $(Ri, Pr, \tau, \Lambda)=(1, 7, 0.01, 2)$. Statistical data of figure \ref{fig:dns_flow_pattern} are presented in table \ref{tab:dns_data} in Appendix \ref{app:dns_data}. DNS is performed such that the flow is fully developed with one-layer convection as a final state during a sufficient time interval for statistics. We use the time-averaged quantities, including the Nusselt number ($\overline{Nu}$), Sherwood number ($\overline{Sh}$), and total flux ratio ($\overline{\gamma}$), where $\overline{\gamma}$ is defined as follows: 
\begin{equation}
\overline{\gamma}=\frac{\overline{\beta_S(\langle w^*S^*\rangle_h-\kappa_S\langle\partial_{z^*} S^*\rangle_h)|_{z^*=0}}}{\overline{\beta_T(\langle w^*T^*\rangle_h-\kappa_T\langle\partial_{z^*} T^*\rangle_h)|_{z^*=0}}}=\Lambda\tau \frac{\overline{Sh}}{\overline{Nu}},
    \label{eq:gamma}
\end{equation}
where the angle brackets  $\langle\cdot\rangle_h$ with subscript `$h$' represent averaging over a horizontal plane, and the overline $\overline{(\cdot)}$ indicates the time average. As shown in figure \ref{fig:dns_flow_pattern}, in general, the fluxes ($\overline{Nu}$, $\overline{Sh}$) increase at a higher value of $Ra$ due to a stronger convection. Notably, there are decreases in $\overline{Nu}$ at $Ra\in[3\times10^4, 4\times10^4]$ and $\overline{Sh}$ at $Ra\in[3\times10^4, 4\times10^4]$ and $Ra\in [1\times10^5, 2\times10^5]$. When $\overline{Sh}$ decreases within $Ra\in[1\times10^5, 2\times10^5]$, $Nu$ still remains increasing, leading to a dramatic decrease in $\overline{\gamma}$ in figure \ref{fig:dns_flow_pattern}(c). Following theoretical predictions by \citet{linden1978diffusive}, the transport in a diffusive DDC flow (without background shear) is dominated by the diffusion process. Thus, for DDC-dominated flow, the value of $\overline{\gamma}$ should be approximately equal to $\sqrt{\tau}=0.1$. However, $\overline{\gamma}$ here is smaller than $\sqrt{\tau}=0.1$. Clearly, to reach the upper bound $\sqrt{\tau}$ of the DDC-dominated regime, it is necessary to conduct DNS with a high $Ra>10^6$, which is the focus of the works of \cite{yang2022layering, li2024double}. In fact, \citet{li2024double} showed that $Ra\geq10^7$ is required to approach the flux ratio $\gamma \approx  \sqrt{\tau}$. Therefore, the SDDC for $Ra\leq 10^6$ in this study is in the shear-influenced regime instead of the DDC-dominated regime, as shear flow can reduce the convective flux. 

To confirm the shear-influenced regime within the present parameter regime, we have computed the contributions of shear production and buoyant energy input to the kinetic energy balance. By taking the inner product between $\boldsymbol{U}$ and \eqref{eq.sddc_nondimensional.momentum} and then taking the volume average ($\langle \cdot \rangle$) over the entire domain, the kinetic energy equation can be determined by
\begin{equation}
\label{eq:energy_balance}
    \frac{1}{2} \frac{\partial}{\partial t} \left\langle \boldsymbol{U}^2 \right\rangle = \sqrt{\frac{Pr}{Ra}} \tau_w U_w + \left\langle (T-\Lambda S)w \right\rangle - \sqrt{\frac{Pr}{Ra}} \left\langle (\nabla \times \boldsymbol{U})^2 \right\rangle = I_w + I_b - D,
\end{equation}
where $I_w$ and $I_b$ are the rates of energy input from the walls \citep{halcrow2008charting} and buoyancy forces \citep{reetz2019turbulent}. In equation \eqref{eq:energy_balance}, $D$ is the viscous dissipation, and $\tau_w$ is the wall-shear stress
\begin{equation}
    \tau_w := 1 + \frac{1}{2L_x L_y} \int_{0}^{L_x} \int_{0}^{L_y} \left( \left.\frac{\partial u}{\partial z} \right|_{z=-0.5} + \left.\frac{\partial u}{\partial z} \right|_{z=0.5}\right) \,dx \,dy.
\label{eq:wallshear_stress}
\end{equation}
After performing a time average of \eqref{eq:energy_balance}, all equilibrium states and statistically steady states should satisfy $\overline{I}_w + \overline{I}_b = \overline{D}$. Our computations indicate that all cases $Ra\leq10^6$ in figure \ref{fig:dns_flow_pattern} have a ratio of $\overline{I}_w/\overline{D}$ higher than 90\% (or $\overline{I}_b/\overline{D}$ less than 10\%), implying that the fluid flow is significantly influenced by shear force. This finding indicates that the present parameter regime is within the shear-influenced regime.

\begin{figure}
\centering
\includegraphics[width=1\linewidth]{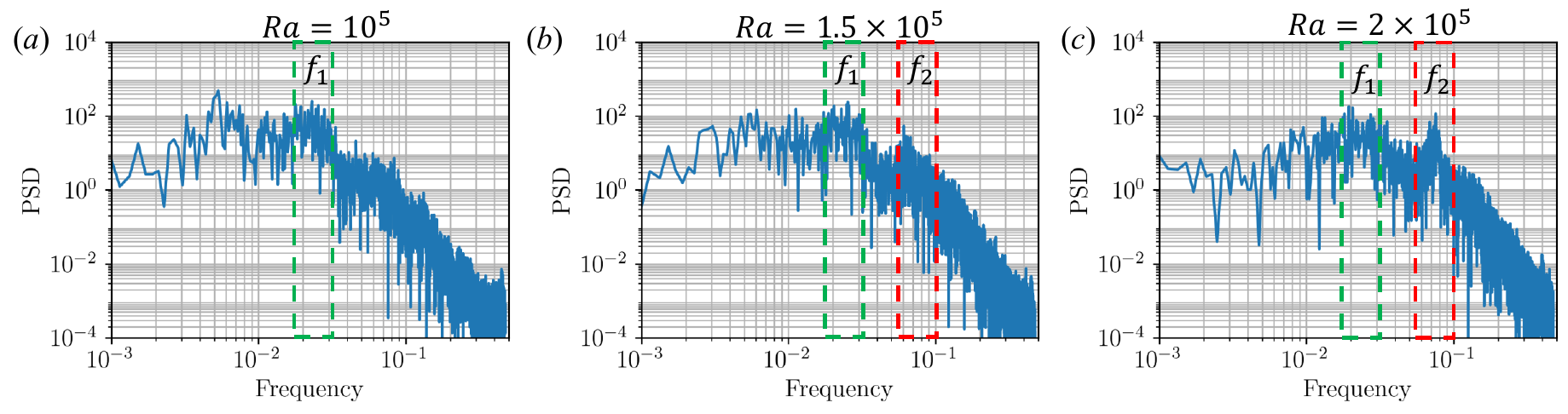}
\caption{Power spectral density (PSD) of $Nu$ for different $Ra$. Green and red boxes of dashed lines indicate the interested zones of frequencies ($f_{1,2}$), representing differently dominant flow structures with and without dominant tilted rolls, respectively, near the onset of the shear-influenced regime.}	
\label{fig:psd}
\end{figure}

Regarding the flow structure behavior, we found that laminar (trivial) flow is dominant for $Ra\leq 10,000$, indicating the laminar flow regime. In the increase of $Ra$, SDDC becomes steady non-trivial flow states and then transitions to unsteady non-trivial flow states with $Ra>30,000$, as shown by the white and gray zones in figure \ref{fig:dns_flow_pattern}(a,b). In the steady flow regime, the flow exhibits tilted rolls. The number of rolls depends on the initial conditions, suggesting the potential coexistence of multiple roll states. When the flow is unsteady at a higher $Ra$, the convective fluxes vary over time, leading to a drop in $\overline{Nu}$ and a decrease in $\overline{Sh}$. Although DNS exhibits chaotic behavior, it still shows instantaneous flow structures resembling tilted rolls, indicating that chaos is influenced by the roll states. As highlighted by the blue zone in figure \ref{fig:dns_flow_pattern}(c), this chaos of tilted rolls is observed until approximately $Ra=1.5\times 10^5$, where a significant decrease in $\overline{\gamma}$ occurs. This decrease of $\overline{\gamma}$ is associated with the reduction and disappearance of roll states, and the flow structures at $Ra=2\times 10^5$ are similar to the final state in figure \ref{fig:dns_sddc}(b). Moreover, this decrease in $\overline{\gamma}$ is associated with the emergence of a new oscillation frequency, as shown by the power spectral density of $Nu$ at various $Ra$ in figure \ref{fig:psd}. At $Ra=10^5$ in figure \ref{fig:psd}(a), we observed $f_{1}$ (green box) less than approximately $1/10^{-1.8}$ or time period greater than $63.1$. At $Ra=1.5\times10^5$ and then $Ra=2\times10^5$ where $\overline{\gamma}$ decreases (figure \ref{fig:dns_flow_pattern}c), another flow response frequency $f_2$ emerges (red boxes), as shown in figure \ref{fig:psd}(b,c).

\section{Equilibrium solutions}\label{sec:equilibrium}

\subsection{Multiple-roll states}\label{sec:convection_roll}
\begin{figure}
\centering
\includegraphics[width=1\linewidth]{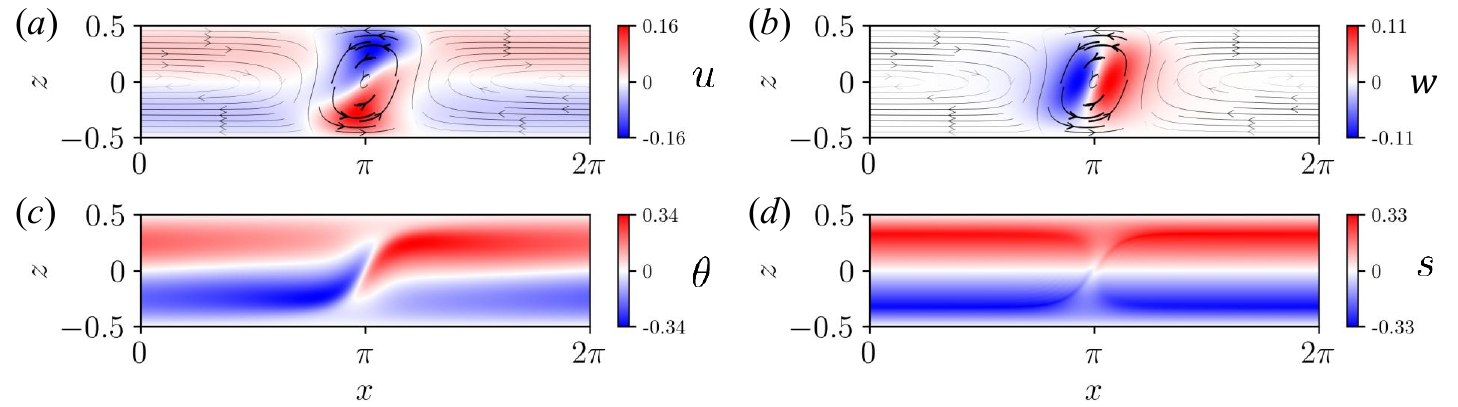}
\caption{Fluctuations of (a) streamwise velocity $u$, (b) vertical velocity $w$, (c) temperature $\theta$, and (d) salinity $s$, illustrating one pair of counter-rotating convection rolls (E1) for SDDC at parameters $(Ra, Ri, Pr, \tau, \Lambda)=(10^5, 1, 7, 0.01, 2)$.}
\label{fig:E1_perturbations}
\end{figure}

Roll states are typical flow structures established by one or multiple counter-rotating convection rolls, which are often observed in natural convection and double-diffusive convection \citep{gao2013transition,liu2022single,zheng2024natural}. When a large-scale background shear flow is introduced to the system, the counter-rotating convection rolls are tilted by the shear flow \citep{zheng2024natural}. Similarly, we found two-dimensional equilibrium solutions of convective roll states for SDDC. As a representative solution, Figure \ref{fig:E1_perturbations} shows a pair of tilted counter-rotating convection rolls (E1) for SDDC with dimensionless parameters $(Ra,Ri,Pr,\tau,\Lambda)=(10^5,1,7,0.01,2)$. This tilted roll state perfectly translates a streamwise length $\lambda_x=L_x=2\pi$ in the horizontal direction, satisfying the trivial symmetry $\langle \pi_z,\pi_x,\tau_x(L_x) \rangle$ of the SDDC system. As described by its name, E1 includes one pair of counter-rotating rolls with different velocity strengths, as demonstrated by figure \ref{fig:E1_perturbations}(a-b). The weak-strength convective roll has a clockwise direction and is dominant in the whole convective layer. To balance momentum, a counterclockwise roll with stronger strength is established in a narrow area at the center of the layer ($x=\pi$). These convection rolls induce mixing in both the temperature and salinity fluctuations, where the positive part lies mostly in the upper half of the layer; see figure \ref{fig:E1_perturbations}(c-d).

\begin{figure}
\centering
\includegraphics[width=1\linewidth]{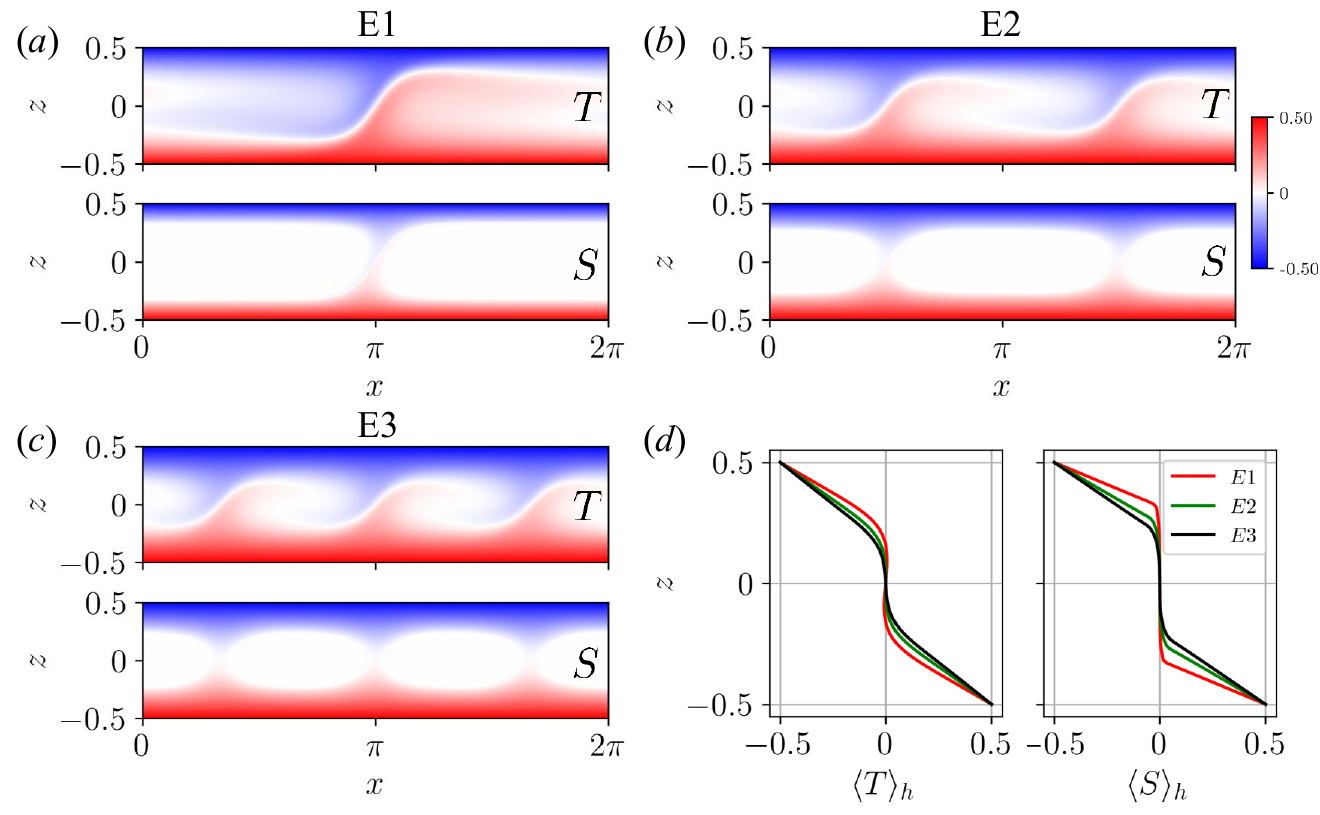}
\caption{ Panels (a)-(c) show total temperature $T$ and total salinity $S$ for equilibrium solutions with (a) one pair (E1), (b) two pairs (E2), and (c) three pairs (E3) of tilted convection rolls for SDDC. These solutions correspond to solutions on the upper branch of the bifurcation diagram shown in Figure \ref{fig:eq_bifurcation}. (d) Profiles of horizontally averaged total temperature $\langle T \rangle_h$ (left) and horizontally averaged total salinity $\langle S\rangle_h$ (right) for E1, E2, and E3 solutions. All panels are associated with the governing parameter $(Ra,Ri,Pr,\tau,\Lambda)=(10^5,1,7,0.01,2)$.}
\label{fig:EQ_totalfields}
\end{figure}

In addition to the E1, we also found co-existence of equilibrium solutions associated with multiple-pair convection rolls. Here, we show solutions with two (E2) and three pairs (E3) of tilted counter-rotating convection rolls as a representation of the multiple roll states. These two solutions satisfy the symmetries $\langle \pi_z,\pi_x,\tau_x(L_x/2) \rangle$ and $\langle \pi_z,\pi_x,\tau_x(L_x/3) \rangle$, respectively. In general, these convective roll states are qualitatively similar to E1. The remarkable difference here is the horizontal translation wavelength of convection rolls with $\lambda_x=L_x/2$ for E2 and $\lambda_x=L_x/3$ for E3. Note that E2 and E3 also satisfy the trivial symmetry $\langle \pi_z,\pi_x,\tau_x(L_x) \rangle$. A qualitatively E3-like equilibrium state for the temperature field with a translation wavelength of $10/3$ was found in vertical convection with a streamwise domain size of $10$ and a cubic background velocity profile \citep{zheng2024natural}. As shown in figure \ref{fig:EQ_totalfields}(a-c), three tilted roll states E1-E3 are illustrated by total temperature and salinity fields for SDDC corresponding to the same dimensionless parameters in figure \ref{fig:E1_perturbations}. Convection rolls generate well-mixed regions indicated by white-colored areas, especially the salinity. Qualitatively, we can see that the area of the well-mixed regions of E1 is the largest. As the number of convection rolls increases, the area of well-mixed regions reduces in both horizontal and vertical directions. This difference of well-mixed region can also be demonstrated by mean profiles of total temperature and salinity fields in figure \ref{fig:EQ_totalfields}(d). The well-mixed regions of the total salinity profile decrease as the number of convection rolls increases, with E1 showing the largest well-mixed regions. The difference in mean profiles between states E2 and E3 is relatively small. Moreover, temperature profiles display a smaller well-mixed region compared to salinity because the diffusivity of temperature is much higher than that of salinity ($\tau=0.01$). The present 2D solutions generate 2D wavy layers near the wall, resembling tunnel structures observed in field measurements through underwater vehicles within ice-shelf cavity \citep[Figure 4]{washam2023direct}.

To understand the bifurcation of equilibrium solutions for SDDC, we performed parametric continuation with the Rayleigh number $Ra$ as a control parameter. We built the bifurcation diagram with response parameters as the convective boundary fluxes of two scalar fields, including the Nusselt number ($Nu$) and the Sherwood number ($Sh$), defined in equation \eqref{eq:Nu_Sh}. The $Nu$ and $Sh$ measure the total flux of both conduction and convection. Both $Nu$ and $Sh$ are independent over $z$ for steady-state solutions (e.g., E1-E3), while they can show dependence over $z$ for certain instantaneous snapshots obtained from DNS.

\begin{figure}
\centering
\includegraphics[width=1\linewidth]{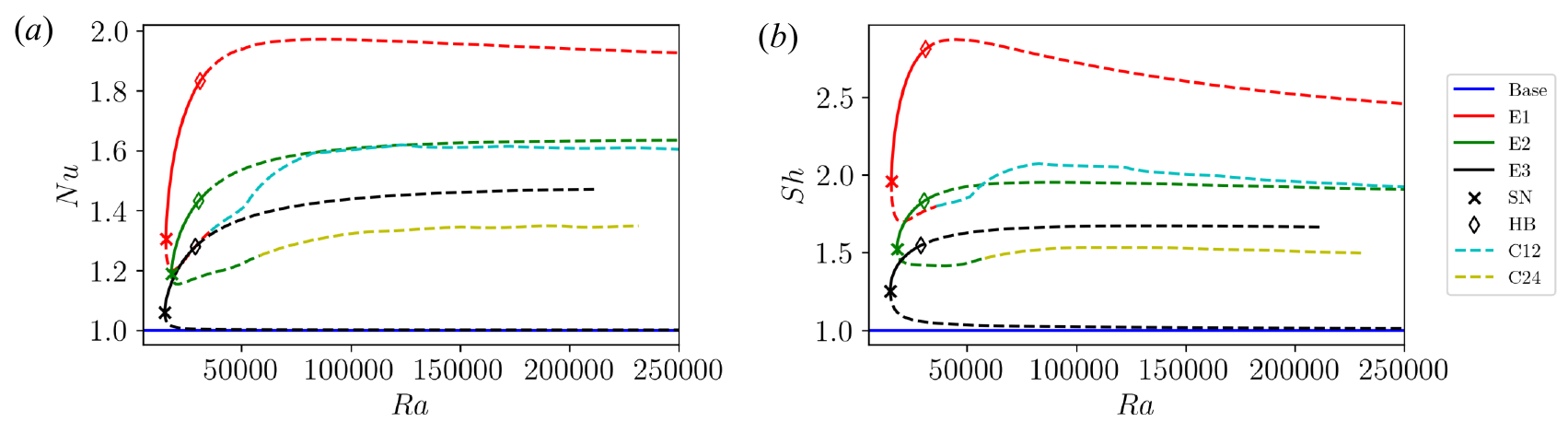}
\caption{Bifurcation diagram of the equilibrium solutions using the (a) Nusselt number ($Nu$) and (b) Sherwood number ($Sh$) as the response parameter and Rayleigh number ($Ra$) as the control parameter. The horizontal blue line denotes the conduction base state associated with $Nu=Sh=1$. The E1-E3 equilibrium solutions in the form of convective roll states are presented by red, green, and black colors, respectively. Cyan and olive colors indicate mixed modes (C12 and C24) of convective roll states. Symbols $\times$ and $\Diamond$ present the saddle-node (SN) bifurcation and Hopf bifurcation (HB) points, respectively. Solid lines indicate stable solutions, while dashed lines indicate unstable solutions.}
\label{fig:eq_bifurcation}
\end{figure}

Figure \ref{fig:eq_bifurcation} displays bifurcation diagrams of equilibrium solutions using $Nu$ and $Sh$ as functions of $Ra$. We start with the conductive base flow, which is stable across the entire range of $Ra$ in our present study. They consistently agree with the linear stability analysis of the base state detailed in Appendix \ref{app:LSA_base_SDDC}, which indicates that the conductive base state is linearly stable within the ranges of $Ra\in[1, 10^7]$ and $Ri\in[0.001,10]$ for fixed parameters $(Pr, \tau, \Lambda)=(7, 0.01, 2)$. This suggests that the nonlinear solutions within SDDC need to be triggered by finite amplitude solutions. This observation is consistent with \citet[\S 4]{li2024double}, which found that SDDC requires an initial disturbance with an amplitude $\delta$ higher than a critical value $1/k_z$ ($k_z$ is the vertical wavenumber of the initial disturbance) to trigger flow behavior different from the conductive base state. This behavior is also similar to plane Couette flow, where the laminar base flow is stable for any Reynolds number \citep{romanov1973stability}, and transition to turbulence needs to be triggered by finite amplitude disturbances; see e.g., \citep{reddy1998stability}.

For convective roll states, $Nu$ and $Sh$ generally decrease with the rise in the number of convection rolls, following the trend of $\text{E1}>\text{E2}>\text{E3}$. As $Ra$ increases, three equilibrium states E1-E3 undergo saddle-node bifurcations, with the stable upper branch and the unstable lower branch solutions near the saddle-node bifurcations. As shown in table \ref{tab:critical_Ra}, the critical values $Ra_\text{SN}$ of the saddle-node bifurcation point for E1-E3 are different. The saddle-node bifurcation for E3 occurs earlier at $Ra_\text{SN}=14,880$ compared with $Ra_\text{SN}=15,482$ for E1 and $Ra_\text{SN}=18,040$ for E2. The stable states on the upper branch are broken into unstable states by Hopf bifurcations, leading to periodic orbits that will be discussed in section \ref{sec:periodic_orbit}. The Hopf bifurcation points for E1-E3 are $Ra_\text{HB}=31069$, $30469$, and $27219$ as listed in table \ref{tab:critical_Ra}. This shows that the Hopf bifurcation appears earlier when the number of convective rolls rises. Note that these values are found by bisection search with an accuracy of $\pm 1$, and bifurcation points for saddle-node and Hopf bifurcations are defined by the transition between stable and unstable states.

\begin{table}
  \begin{center}
\def~{\hphantom{0}}
  \begin{tabular}{lcc}
      State  & $Ra_\text{SN}$ &   $Ra_\text{HB}$ \\[3pt]
       E1   & 15,482 & 31,069  \\
       E2   & 18,040 & 30,469  \\
       E3  & 14,880 & 27,219  \\
  \end{tabular}
  \caption{Critical Rayleigh numbers of the saddle-node ($Ra_{\text{SN}}$) and Hopf bifurcations ($Ra_{\text{HB}}$) for the multiple-roll states E1-E3.}
  \label{tab:critical_Ra}
  \end{center}
\end{table}

\begin{figure}
\centering
\includegraphics[width=1\linewidth]{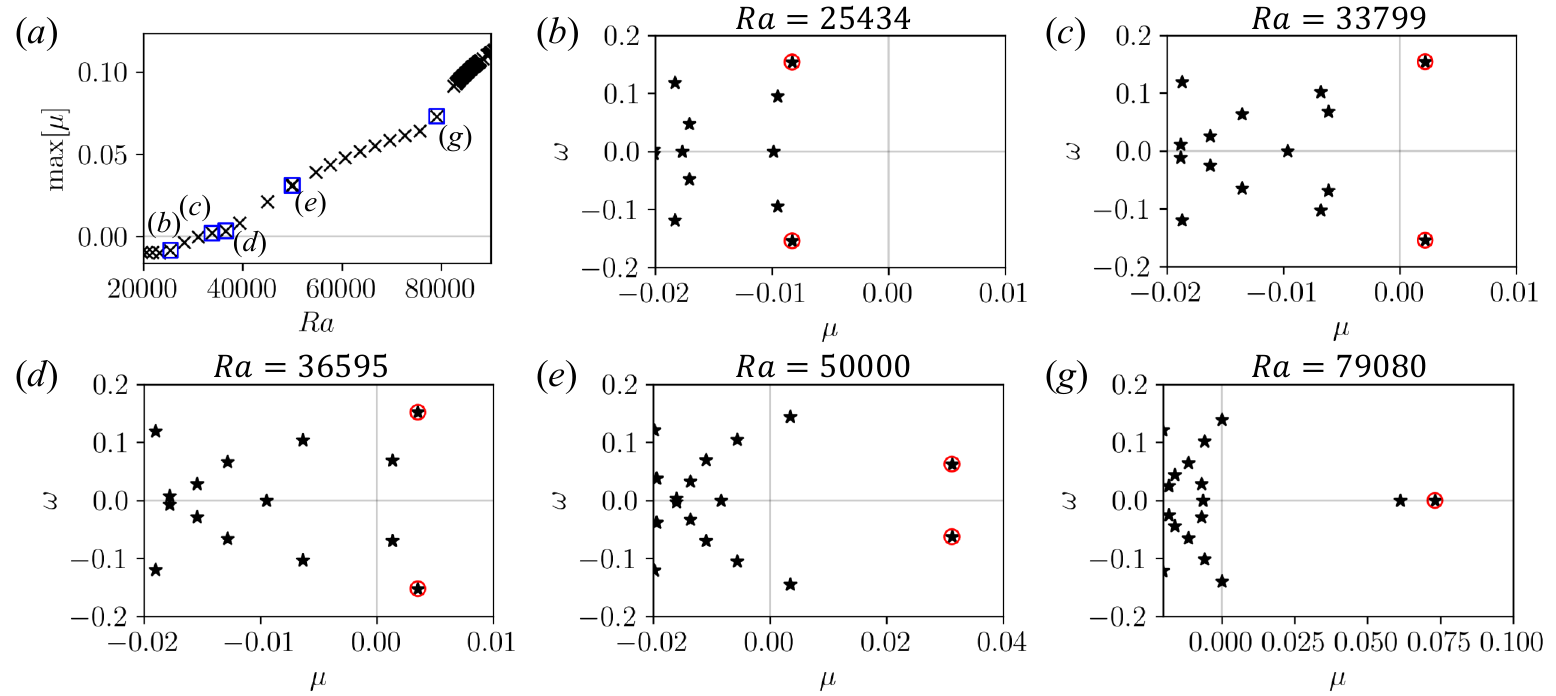}
\caption{Evolution of leading eigenvalues ($\lambda=\mu\pm\text{i}\omega$ with $\mu,\omega\in\mathbb{R}$ and $\text{i}=\sqrt{-1}$) on the upper branch of E1 near Hopf bifurcation. (a) The largest real part of the eigenvalues $\max[\mu]$. Panels (b–g) show the distribution of eigenvalues for five different $Ra$ corresponding to blue squares in panel (a). In panels (b-g), red circles indicate the first leading eigenvalue with the largest real part, which can be a pair of complex conjugate values or a real value.}	
\label{fig:eigen_E1}
\end{figure}

Figure \ref{fig:eigen_E1} displays the development of the spectral eigenvalues ($\lambda=\mu\pm\text{i}\omega$, where $\text{i}=\sqrt{-1}$ is the imaginary unit) of E1 for SDDC, corresponding to the upper branch near the Hopf bifurcation point. We use this representative case to analyze the transition from the stable state to the unstable state of convective roll states. The largest real part of the eigenvalues ($\max[\mu]$) is shown in figure \ref{fig:eigen_E1}(a) to identify the instability. The real part of the leading eigenvalue crosses zero at $Ra_\text{HB}=31,069$ to become positive, indicating an unstable state. The distributions of eigenvalues for E1 at five different $Ra$ are shown in figure \ref{fig:eigen_E1}(b–g), which correspond to the blue squares in figure \ref{fig:eigen_E1}(a). At $Ra=25,434$, all leading eigenvalues are negative as shown in figure \ref{fig:eigen_E1}(b), indicating that the convective roll state is stable. The leading eigenvalues is a pair of complex conjugate ($\lambda_{1,2}$) with an angular frequency $\omega_{\lambda_{1,2}}\approx 0.154$ at $Ra=25,434$ (figure \ref{fig:eigen_E1}b). When $Ra>Ra_\text{HB}$, the leading complex conjugate eigenvalues cross zero as illustrated at $Ra=33,799$ in figure \ref{fig:eigen_E1}(c), resulting in an unstable state characterized by oscillation. It suggests a Hopf bifurcation with a period associated with the angular frequency $T_{po}=2\pi/\omega\approx40$. There is a secondary leading unstable eigenvalue that has a lower angular frequency $\omega_{\lambda_{3,4}}\approx0.062$ that appears at $Ra=36,595$; see figure \ref{fig:eigen_E1}(d). The secondary unstable complex eigenvalues then become the leading eigenvalues of E1 at $Ra=50,000$ (figure \ref{fig:eigen_E1}(e)). The change of the leading complex conjugate of the eigenvalue implies a different dominant oscillatory instability for E1. At $Ra=79,080$ in figure \ref{fig:eigen_E1}(g), the unstable eigenvalues become real with a larger growth rate, which indicates that E1 becomes more unstable as $Ra$ rises.

On the other hand, the lower branches of the convective roll state show different behaviors. In particular, the lower branch of E3, which has the smallest ($Nu,Sh$), shows a subcritical bifurcation in which it would bifurcate from the trivial solution (base state). As shown in figure \ref{fig:eq_bifurcation}, E3 undergoes unstable states on the lower branch approaching $Nu=Sh=1$ that are nearly superimposed on the base state. On the lower branches of E1 and E2, $Nu$ and $Sh$ reduce immediately after the saddle-node bifurcation and then grow, leading to mixed modes that will be discussed in more detail in subsection \ref{sec:mixed_modes}. 

\subsection{Mixed modes}\label{sec:mixed_modes}
\begin{figure}
\centering
\includegraphics[width=0.7\linewidth]{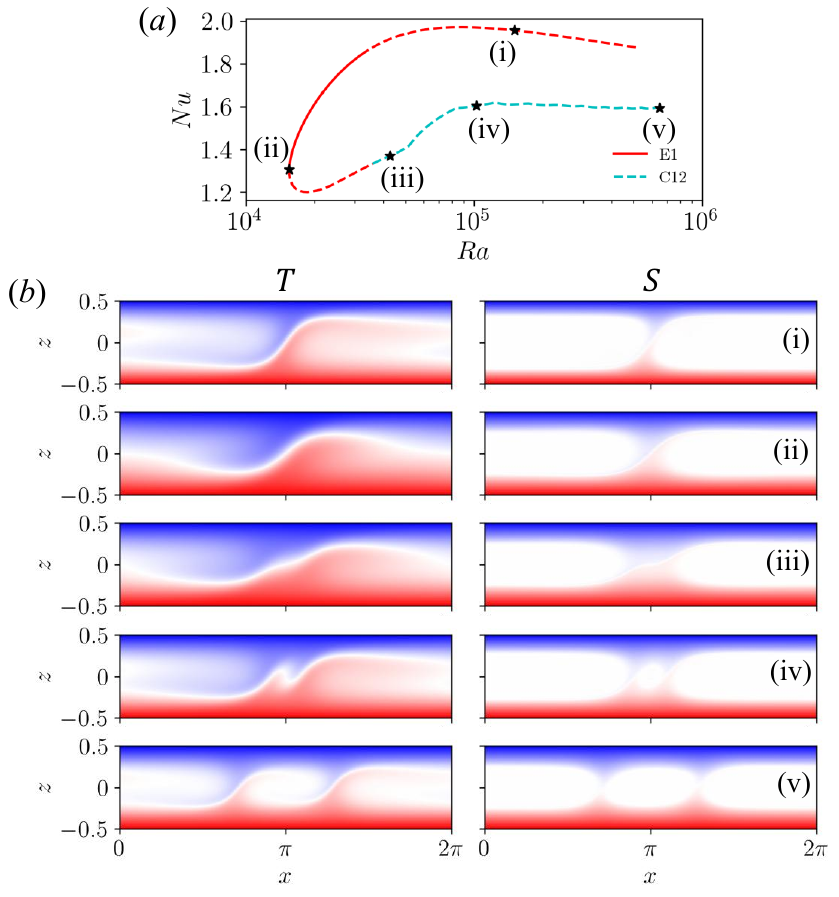}
\caption{(a) Bifurcation diagram of convective roll state E1 (red) and mixed mode C12 (cyan) showing $Nu$ as a function of $Ra$. Bifurcation is plotted in an extended $Ra$ range and logarithmic scale. Solid lines indicate stable solutions, while dashed lines indicate unstable solutions. (b) Development of E1 and C12 illustrated by distributions of total temperature $T$ (left) and total salinity $S$ (right) fields at five different points corresponding to star symbols in panel (a).}	
\label{fig:mixed_modes}
\end{figure}

In this section, we discuss mixed modes of convective roll states. A convective roll state has a horizontal translation symmetry $\tau(\lambda_x)$ of horizontal wavelength $\lambda_x=L_x/n$, where $n$ represents the number of pairs of convection rolls. For example, E3 has a horizontal translation symmetry $\tau(\lambda_x)$ with $\lambda_x=L_x/3$. This symmetry operator is ignored in the mixed modes, leading to trivial symmetry. In this subsection, two mixed modes, named C12 and C24, were identified. As shown by cyan and yellow colors in figure \ref{fig:eq_bifurcation}, C12 connects the lower branch of  E1 solution and the upper branch of E2 solution, while C24 signifies the connection between E2 and E4 solutions. Note that E4 is a convective roll state with a horizontal translation symmetry of a wavelength $\lambda_x=L_x/4$ that is not shown in this subsection for brevity but will be discussed in Section \ref{sec:parameter_continuation}. Throughout this paper, the transition point from convective roll states to mixed modes is determined by the first addition of unstable complex conjugate eigenvalues along the lower branches in the bifurcation diagram in figure \ref{fig:eq_bifurcation}. We use this criterion because the mixed mode will connect with the upper branch equilibrium solutions after the Hopf bifurcation, where we expect unstable complex conjugate eigenvalues. Here, C12 and C24 in figure \ref{fig:eq_bifurcation} show zig-zag curves. When $Ra$ increases, there is a change in the number of unstable leading eigenvalues on C12, which can lead to an additional bifurcation.

We then take a closer look at the solution profile of mixed mode. Figure \ref{fig:mixed_modes}(a) shows the bifurcation diagram of the E1 solution and the mixed mode C12 over an extended $Ra$ range using a logarithmic scale, where we select five points along this curve to display their solution profiles. In figure \ref{fig:mixed_modes}(b), we plot the corresponding total temperature and salinity fields of convective roll states E1 and the mixed mode C12. Points (i) for $Ra=1.5\times 10^5$ and (ii) for $Ra_\text{SN}=15,482$ represent the upper branch and the saddle-node bifurcation point of E1, respectively. When $Ra$ increases on the upper branch, the SDDC system will be increasingly influenced by convection, which governs the mixing process. Clearly, the area of the well-mixed region of point (i), corresponding to the upper branch, is the largest compared with the remaining points. At the saddle-node bifurcation point, there is a reduction in the well-mixed region. Moving along the lower branch, a new flow structure referred to as the mixed-mode C12 appears. Points (iii–v) denote the variation of mixed-mode C12 under $Ra$. At $Ra=42,810$ (point (iii)), the leading unstable eigenvalue is complex. The small roll at $x=\pi$ is about to emerge, although it is still not qualitatively visible in both the total temperature and salinity fields. With higher $Ra$, the new convection rolls are evident, leading to new well-mixed regions, as shown in snapshots (iv) for $Ra=102,154$ and (v) for $Ra=647,208$. At a representation of high $Ra$, point (v) demonstrates the fundamental properties of two pairs of convection rolls in the mixed mode, which are similar to the E2 solution shown in figure \ref{fig:EQ_totalfields}(b). 

\section{Periodic orbits}\label{sec:periodic_orbit}
\begin{figure}
\centering
\includegraphics[width=1\linewidth]{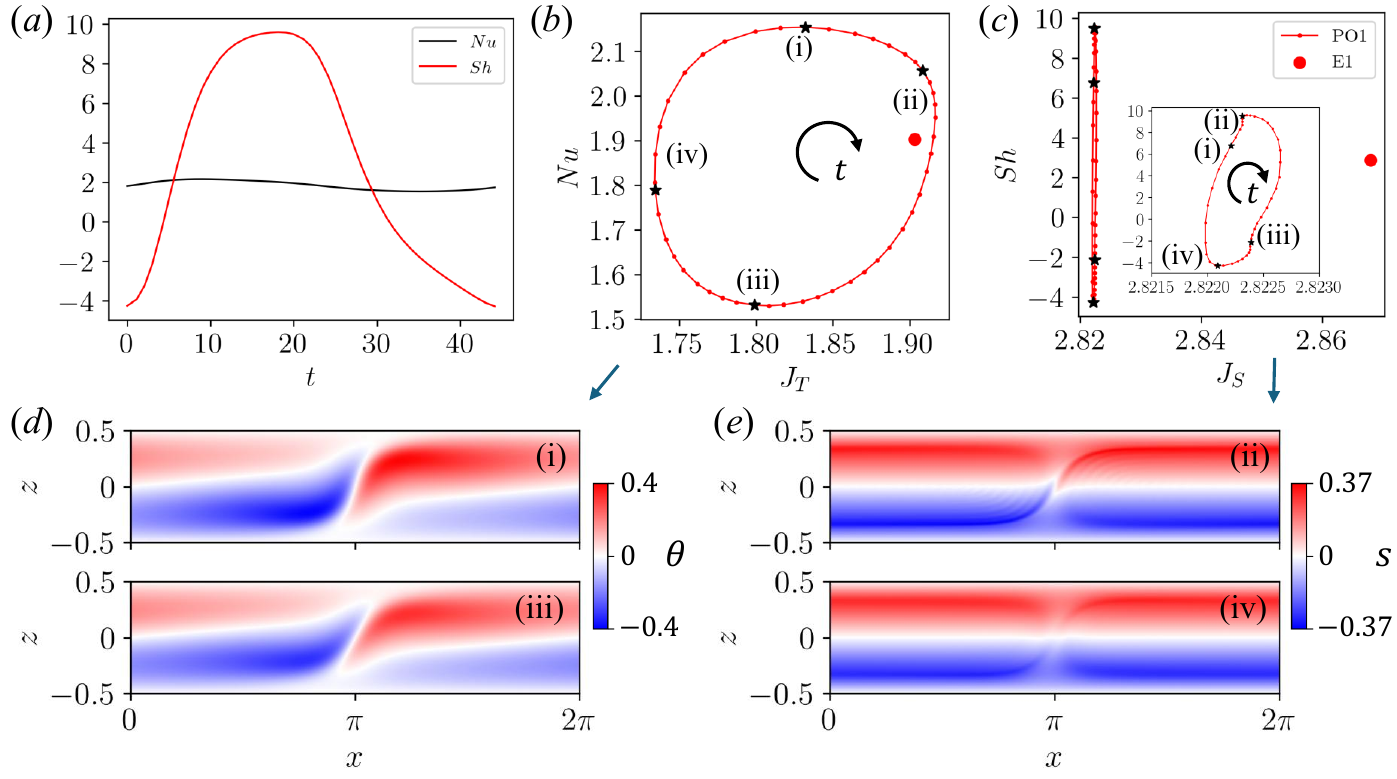}
\caption{Representative properties of periodic orbit PO1 for SDDC at $Ra=40000$. (a) Time history of $Nu$ and $Sh$ numbers of PO1's DNS. (b-c) Periodic orbits in $J_T-Nu$ and $J_S-Sh$ phase spaces with corresponding equilibrium states (star symbol). The arrow denotes the time evolution direction consistent with panel (a). Panel (d,e) shows snapshots of temperature ($\theta$) and salinity ($s$) fluctuations.} 	
\label{fig:po1}
\end{figure}

\begin{figure}
    \centering
    \includegraphics[width=1\linewidth]{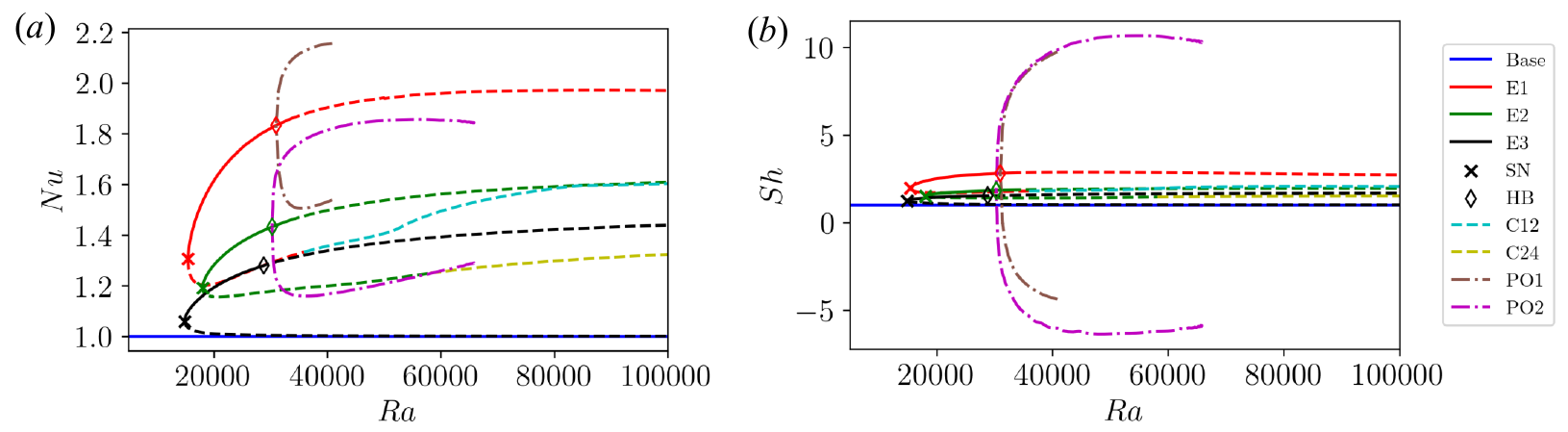}
    \caption{Bifurcation diagram of equilibrium states and two periodic orbits showing (a) the Nusselt number ($Nu$) and (b) the Sherwood number ($Sh$) as functions of the Rayleigh number $Ra$. The conduction base state is denoted by the blue horizontal line with $Nu=Sh=1$. Convective roll states E1-E3 are presented by red, green, and black colors. Cyan and olive colors indicate mixed modes (C12 and C24) of convective roll states. Solid and dashed lines indicate stable and unstable equilibrium states, respectively, and we do not compute the stability of periodic orbits. Symbols $\Diamond$ and $\times$ present the Hopf bifurcation (HB) and saddle-node (SN) points, respectively. Periodic orbits PO1 and PO2 are respectively presented by brown and purple dash-dot lines, which are illustrated by maximal and minimal values of $Nu$ and $Sh$ during a single period.}
    \label{fig:EQPO_bifurcation_diagram}
\end{figure}
\begin{figure}
    \centering
    \includegraphics[width=0.5\linewidth]{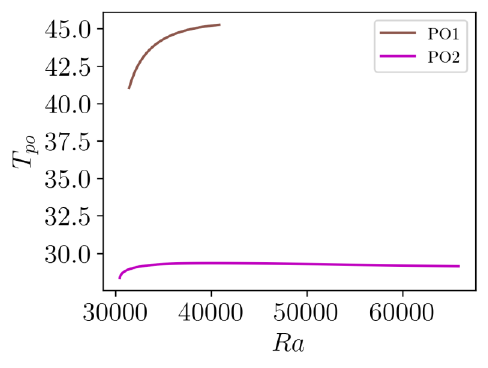}
    \caption{Time period $T_{po}$ of periodic orbits PO1 (brown) and PO2 (purple) as functions of $Ra$.}
    \label{fig:PO_period}
\end{figure}

As discussed in section \ref{sec:equilibrium}, Hopf bifurcations are formed on the upper branch of convective roll states, where periodic orbits can be generated. We found two periodic orbits named PO1 and PO2, implying bifurcations from the equilibrium states E1 and E2, respectively. Here, we do not analyze the stability of PO1 and PO2 by Floquet theory, and we only characterize the solution behavior of these periodic orbits.

Figure \ref{fig:po1} shows the main properties of the periodic orbit PO1, represented by the PO1 state at $Ra=40000$. Time histories of the $Nu$ and $Sh$ numbers over a single time period $T_{po}=45.21$ are plotted in figure \ref{fig:po1}(a), indicating that there is a phase difference between $Nu$ and $Sh$. In order to clarify correlations between conduction and convection, we plot convective boundary fluxes and corresponding conductive boundary fluxes in dynamic space for both temperature and salinity during a simple period in figure \ref{fig:po1}(b-c). Here, conductive boundary fluxes for total temperature and salinity are defined as follows \citep{reetz2019turbulent}: 
\begin{subequations}
\begin{align}
    J_T=&\frac{-\langle \partial_{z^*} T^* \rangle_h|_{z^*=-0.5H}}{\Delta T^* H^{-1}}=-\langle \partial_z T\rangle_h|_{z=-0.5},\\
    J_S=&\frac{-\langle \partial_{z^*} S^* \rangle_h|_{z^*=-0.5H}}{\Delta T^* H^{-1}}=-\langle \partial_z S\rangle_h|_{z=-0.5}.
    \end{align}
\end{subequations}
For temperature, the $J_T-Nu$ curve of orbit PO1 is smooth, and the equilibrium state E1 is located inside but close to the right-hand pole of the $J_T-Nu$ curve of PO1. On the other hand, the $J_S-Sh$ orbit, which is bean-shaped, indicates a sharp transition at the maximum peak of $Sh$. Interestingly, the $J_S-Sh$ of the equilibrium state E1 is located outside the $J_S-Sh$ curve associated with PO1. In periodic orbit PO1, the change of $Nu$ and $J_T$ over time is moderate ($Nu_{max}-Nu_{min}<0.6$) or even small ($J_{T,max}-J_{T,min}<0.2$), while $Sh$ undergoes a higher range and even negative values (see figure \ref{fig:po1}(c)). The snapshots of temperature and salinity for the minimum and maximum peaks over the orbits are displayed in figure \ref{fig:po1}(d-e), where fluctuation fields are used for visualization due to small value differences. Temperature fluctuations at the two opposite peaks are almost similar with respect to their distributions. Snapshot (i) shows the largest amplitude of temperature fluctuations during a time period. The temperature fluctuation in snapshot (iii) is slightly weaker, but it reduces the inclination angle of the convection roll near $x=\pi$. For salinity, the difference between the maximum and minimum peaks is clearly displayed in figure \ref{fig:po1}(e). The maximum peak shown by snapshot (ii) exhibits a sharp inflection at $x=\pi$. At the minimum peak, convective flux is less than conductive flux, leading to negative $Sh$. As a result, the sharp inflections of the salinity are more blurred, as shown by snapshot (iv) of figure \ref{fig:po1}(e). In general, PO2 has similar oscillation properties to PO1; thus, we do not show the solution behavior of PO2 here.

Figure \ref{fig:EQPO_bifurcation_diagram} shows the bifurcation diagram of periodic orbits along with the equilibrium solutions of convective roll states discussed in the previous section \ref{sec:equilibrium}. To illustrate the change of quantities during the periodic orbits, we plot the maximum and minimum peaks of convective fluxes ($Nu,Sh$) during a single time period. $Nu$ and $Sh$ of the E1-E3 solutions almost lie in the middle between the maximum and minimum peaks of PO1 and PO2. Interestingly, bifurcations of convective roll states (E1-E3) and periodic orbits (PO1 and PO2) using $Nu$ are above the unit value of the base state, while $Sh$ for PO1 and PO2 can show negative values. Moreover, $Sh$ during periodic orbits PO1 and PO2 shows a much larger amplitude than that of the equilibrium solutions E1-E3 and even reaches negative values. These behaviors match transitional stages in the statistics of ($Nu$, $Sh$) over time observed by DNS; see e.g., \citet[Figure 4(a,b)]{yang2022layering} and \citet[Figure 3(b,c)]{li2024double}. In addition to the change of convective fluxes during periodic orbits, variations of time periods $T_{po}$ under the change of $Ra$ are displayed in figure \ref{fig:PO_period}. Here, PO1 has a larger time period compared to PO2.  As $Ra$ increases, the period of PO2 quickly increases and then remains almost constant. In fact, the time period of PO1 near the Hopf bifurcation point is close to the predicted period $T=2\pi/\omega$ using the angular frequency $\omega$ obtained from the leading eigenvalues via the linear stability analysis for equilibrium states in figure \ref{fig:eigen_E1}. As implied in figures \ref{fig:EQPO_bifurcation_diagram} and \ref{fig:PO_period}, PO1 and PO2 do not converge with much higher $Ra$.

\begin{figure}
\centering
\includegraphics[width=1\linewidth]{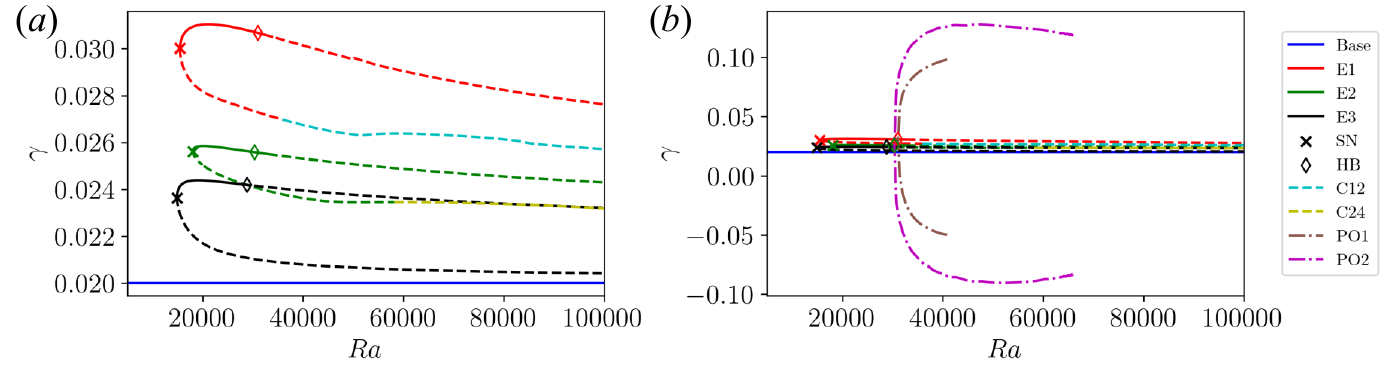}
\caption{Flux ratio $\gamma=\Lambda\tau Sh/Nu$ as functions of $Ra$ (a) without periodic orbits and (b) with periodic orbits. All lines and symbols have the same meaning as figure \ref{fig:EQPO_bifurcation_diagram}. Periodic orbits are illustrated by maximal and minimal values of instantaneous $\gamma$ during a single period.} 
\label{fig:flux_ratio}
\end{figure}

Figure \ref{fig:flux_ratio} displays the flux ratio $\gamma=\Lambda\tau Sh/Nu$ measuring the ratio of density fluxes caused by salinity and temperature. Following theoretical predictions by \citet{linden1978diffusive}, the transport in a diffusive DDC flow is dominated by the diffusion process when the flux ratio $\gamma$ should be approximately equal to $\sqrt{\tau}=0.1$. As shown in figure \ref{fig:flux_ratio}(a), the flux ratio of all the convective roll states is much smaller than $\sqrt{\tau}=0.1$ and close to $\gamma=\Lambda\tau=0.02$, corresponding to the conduction base state associated with $Nu=Sh=1$. The flux ratio decreases with the increase of $Ra$ and the number of convection rolls. The highest values of $\gamma$ are exhibited on the upper branches, where they remain stable. The flux ratios of these convection roll solutions in figure \ref{fig:flux_ratio}(a) are within $\gamma \leq 0.032$, indicating that these solutions are within the shear-influenced regime \citep[Figure 9]{li2024double}. An obvious difference is observed in the flux ratio of periodic orbits in figure \ref{fig:flux_ratio}(b), which also includes the maximal and minimal flux ratios of PO1 and PO2 over one period. Periodic orbit solutions can reach large $\gamma$ instantaneously, especially PO2, whose $\gamma$ maximized over one period can exceed $\sqrt{\tau}=0.1$.  Therefore, SDDC has a transition from steady shear-influenced convection rolls to periodic orbits that includes an instantaneous DDC-dominated process.

\begin{figure}
\centering
\includegraphics[width=1\linewidth]{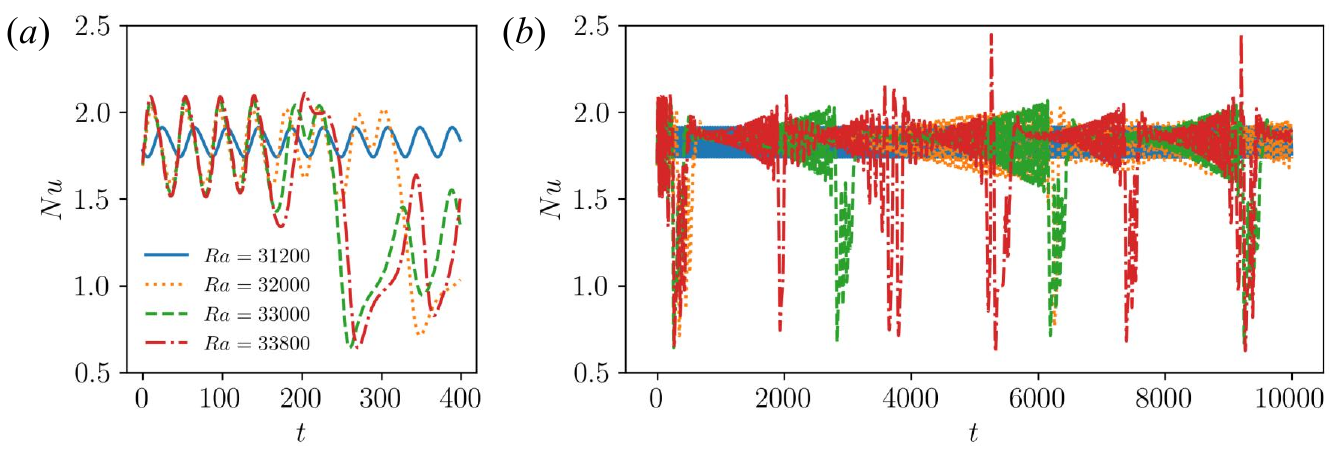}
\caption{DNS obtained from the perturbed PO1 for various $Ra$ near the Hopf bifurcation. A perturbation with a magnitude of $10^{-4}$ is added to PO1 as an initial disturbance. Panel (a) shows the early development of DNS in panel (b).}	
\label{fig:PO_becomes_chaos}
\end{figure}

By bifurcation theory, periodic orbits near the supercritical Hopf bifurcation point (as implied by figure \ref{fig:EQPO_bifurcation_diagram}) should be stable. With an increase in $Ra$, they then become unstable. Although we do not analyze the stability of periodic orbits through Floquet theory, we can quickly check this behavior by numerical simulations. We use a state snapshot during the periodic orbit PO1, perturbed by random noise with a magnitude of $10^{-4}$ in each field, as the initial condition for DNS. As shown in figure \ref{fig:PO_becomes_chaos}(a), the DNS of SDDC at $Ra=31200$ (near the Hopf bifurcation point of PO1) shows stable periods over time, while the DNS for $Ra=32000$, 33000, and 33800 exhibits trajectories moving away from the periodic solution PO1 after three periods. Figure \ref{fig:PO_becomes_chaos}(b) displays the DNS response over a longer time. The resulting observations for $Ra=32000$, 33000, and 33800 are similar to the bursting behavior \citep{viswanath2007recurrent}, in which $Nu$ bursts occur over a long time. Here, the DNS frequently converges towards E1 and slowly moves away in E1's unstable manifold, ultimately approaching  PO1. Because PO1 is unstable, the DNS rapidly progresses to the stable manifold of E2 (or PO2) before returning to E1's stable manifold. Bursting time intervals tend to be faster in the rise of $Ra$ (figure \ref{fig:PO_becomes_chaos}b). As $Ra$ increases further, the DNS exhibits more chaotic trajectories and subsequently develops into chaos.

\begin{figure}
\centering
\includegraphics[width=1\linewidth]{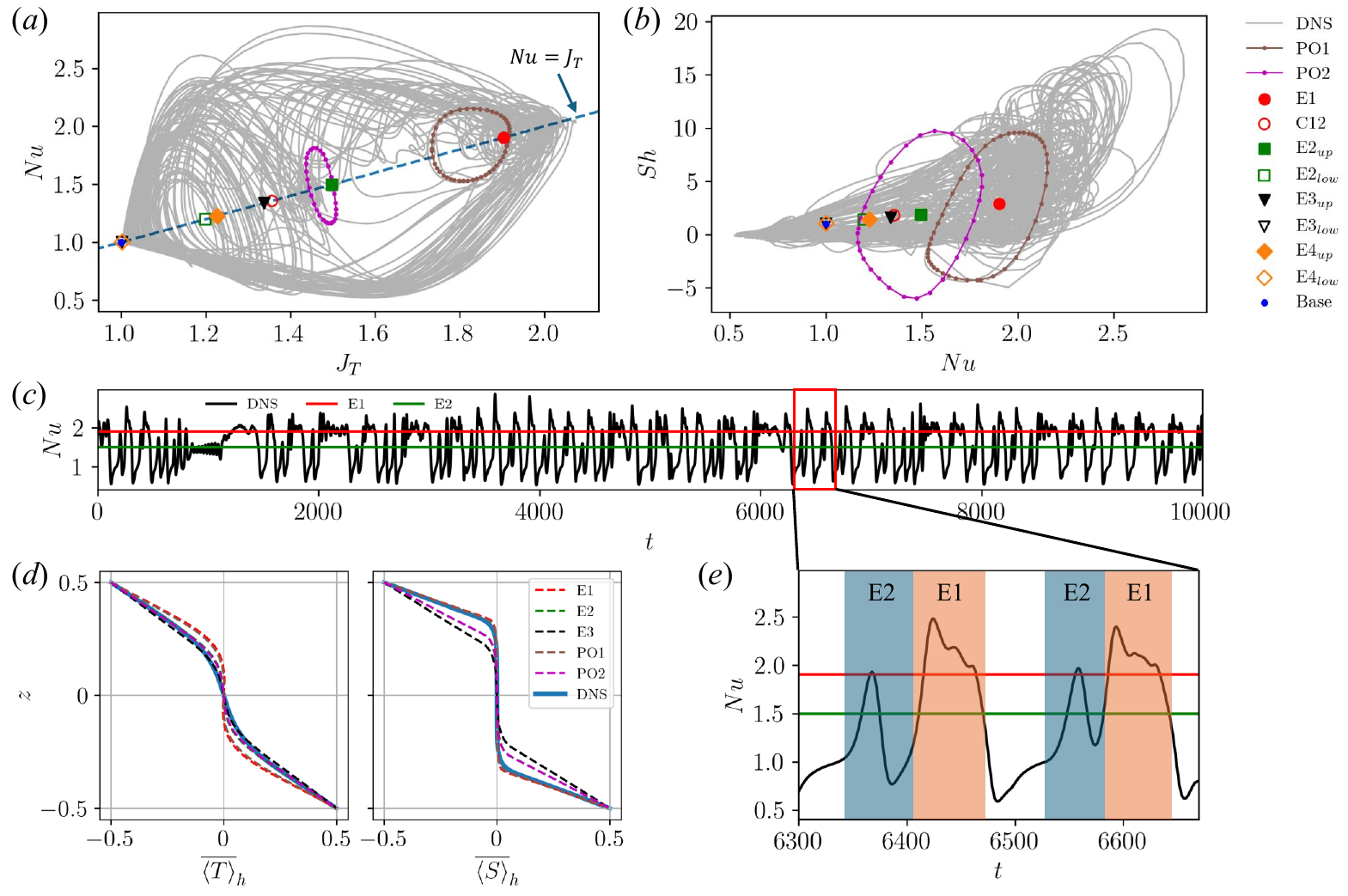}
\caption{Properties of SDDC at $Ra=40000$: (a,b) Phase portraits $J_T-Nu$ and $Nu-Sh$ of DNS. Equilibrium states (E1, C12, $\text{E2}_{up}$, $\text{E2}_{low}$, $\text{E3}_{up}$, $\text{E3}_{low}$, $\text{E4}_{up}$, $\text{E4}_{low}$) and periodic orbits (PO1, PO2) are added, where `up' and `low' subscripts indicate the corresponding upper branch and lower branch equilibrium solutions, respectively. (c) Time history of $Nu$ obtained from DNS. (d) Comparison of mean profiles (averaged over horizontal direction and time) of temperature (left) and salinity (right) among DNS, equilibrium solutions, and periodic orbits. (e) Zoom in on panel (c). Orange and blue boxes denote instantaneous flow behavior similar to E1 and E2, respectively. Horizontal lines in panels (c) and (e) indicate $Nu$ of the equilibrium solutions E1 (red) and E2 (green), respectively. See supplementary movie 2.}	
\label{fig:phase_portrait_Ra40000}
\end{figure}

Figure \ref{fig:phase_portrait_Ra40000} shows the main properties of chaotic dynamics represented by SDDC with dimensionless parameters $(Ra,Ri,Pr,\tau,\Lambda)=(4\times10^4,1,7,0.01,2)$. Figures \ref{fig:phase_portrait_Ra40000}(a,b) display the phase portraits $J_T-Nu$ and $Nu-Sh$ of DNS as the background under all the equilibrium states and periodic orbits. The initial disturbance is the perturbed PO1. All equilibrium states lie on the line of the function $Nu=J_T$, and all of these states in figures \ref{fig:phase_portrait_Ra40000}(a,b) are unstable at these parameters. The response of DNS over time shows chaotic behavior but visits the neighborhoods of unstable equilibrium states and periodic orbits. Figure \ref{fig:phase_portrait_Ra40000}(c) displays $Nu$ over time obtained from DNS. Mean profiles of the total temperature and salinity from DNS (averaged over both time and horizontal directions) are compared with those of equilibrium solutions and periodic orbits in figure \ref{fig:phase_portrait_Ra40000}(d). Mean profiles of PO1 and PO2 almost overlap with the corresponding equilibrium solutions E1 and E2. The response of DNS for total temperature exhibits a mean profile close to E2 (or PO2) and E3, while it is close to E1 (or PO1) for total salinity. It indicates that the mean characteristics of SDDC over a long time are dominated by E1 (or PO1) for salinity and E2 (or PO2) for temperature. Figure \ref{fig:phase_portrait_Ra40000}(e) illustrates the chaotic switching behavior of SDDC through the observation of instantaneous E1- and E2-like flow structures when DNS visits the stable and unstable manifolds of ECS (E1, PO1, E2, and PO2). A similar phenomenon has been widely observed in existing analyses of ECS and DNS, which highlight the importance of unstable ECS in determining the statistics of chaotic solutions \citep{waleffe2009exact,reetz2019turbulent,wagner2022exact,zheng2024natural,zheng2024natural2,engel2025search}.

\section{Continuation of other parameters}\label{sec:parameter_continuation}

\begin{figure}
\centering
\includegraphics[width=1\linewidth]{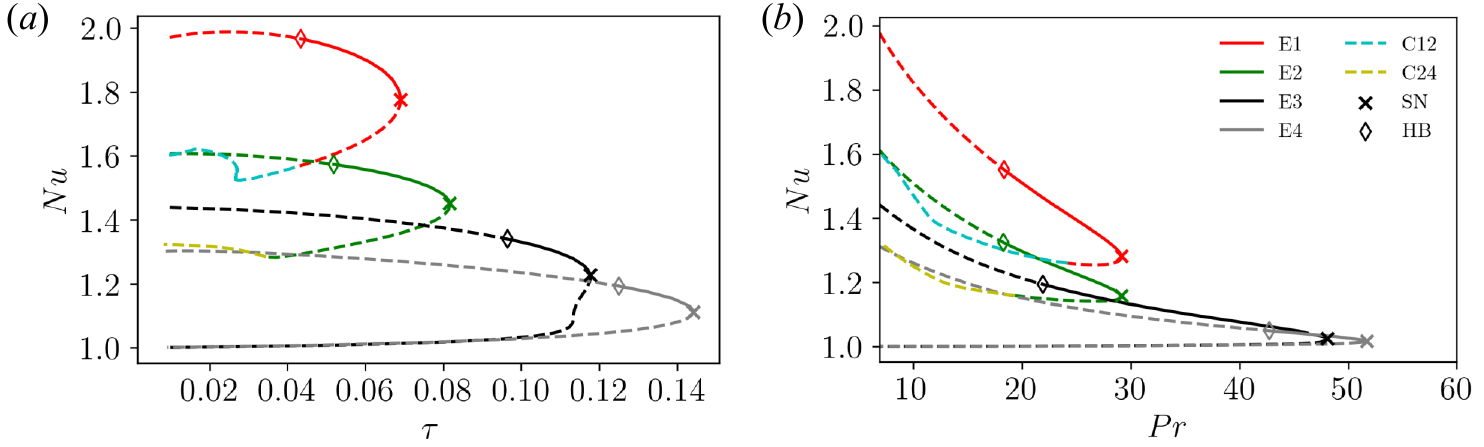}
\caption{Continuation of convective roll states E1-E4 over (a) the diffusivity ratio $\tau$ with fixed parameters $(Ra, \Lambda, Pr, Ri)=(10^5, 2, 7, 1)$ and (b) the Prandtl number $Pr$ with fixed parameters  $(Ra, \Lambda, \tau, Ri)=(10^5, 2, 0.01, 1)$. The change of continuations is presented by the Nusselt number $Nu$. The solid and dashed lines denote the stable and unstable solutions, respectively. Symbols $\times$ and $\Diamond$ present the saddle-node (SN) bifurcation and Hopf bifurcation (HB) points, respectively.}
\label{fig:tau_continuation}
\end{figure}

The diffusivity ratio $\tau$ and Prandtl number $Pr$ represent different media of double-diffusive convection; see e.g., \citet[Figure 12.1]{radko2013double}. To investigate their effects on convective roll states, we build bifurcation diagrams for these parameters, as shown in figure \ref{fig:tau_continuation}. Figure \ref{fig:tau_continuation}(a) shows the $Nu$ for convective roll states E1-E4 as functions of $\tau$ with fixed dimensionless parameters $(Ra, \Lambda, Pr, Ri)=(10^5, 2, 7, 1)$. In general, E1-E4 exhibit subcritical behavior, which undergoes the Hopf and saddle-node bifurcations like the $Ra$-continuation in figure \ref{fig:eq_bifurcation}. In particular, upper branch solutions near the saddle-node bifurcation points are stable. Lower branches of E1 and E2 lead to the corresponding mixed modes C12 and C24 discussed in section \ref{sec:mixed_modes}, and they are connected to the upper branch solutions of E2 and E4, respectively. Near the saddle-node bifurcation points, $Nu$ on the lower branch of E3 dramatically drops before decreasing gradually, but this does not occur for the other convective roll states E1, E2, and E4. With the rise in the number of convection rolls, the saddle-node bifurcation point occurs at a higher $\tau$ with decreasing flux. Specifically, the saddle-node bifurcation appears at $\tau=0.069$, $0.082$, $0.117$, and $0.144$ for E1, E2, E3, and E4 solutions, respectively. However, E1-E4 cannot continuously connect to a non-trivial solution at $\tau=1$ associated with the single component flow, e.g., the stably stratified PCF. Also, we tested DNS for $\tau=1$ with an initial condition as the upper branch E4 solution at $\tau=0.14$, which shows that the flow will decay to the conductive (base) state. 

Figure \ref{fig:tau_continuation}(b) displays the $Nu$ as functions of $Pr$ for convective roll states E1-E4 at $(Ra, \Lambda, \tau, Ri)=(10^5, 2, 0.01, 1)$. Similar to the diffusivity ratio $\tau$, the convective roll states exhibit subcritical behavior, which undergoes Hopf and saddle-node bifurcations under $Pr$ continuation. With the rise of the number of convection rolls, the saddle-node bifurcation point appears at a higher $Pr$ with a sharper fold. Additionally, a decrease in $Nu$ is observed as $Pr$ increases. At the same parameters of continuation in figure \ref{fig:tau_continuation}(b), we use the solution of E3 at $Pr=40$ as the initial condition to conduct DNS at $Pr=100$. The resulting DNS also decays to the conductive (base) state as the final state.

The subcritical behavior of the Prandtl number continuation in Figure \ref{fig:tau_continuation}(b) shows a fundamental difference compared to the thermohaline shear instability \citep{radko2016thermohaline} induced by a sinusoidal background flow in a vertically periodic domain. Such thermohaline-shear instability remains at the infinite Prandtl number limit; see Appendix \ref{app:LAS_stokes_model} comparing the growth rate at a finite Prandtl number \citep[Figure 10b]{radko2016thermohaline} and the infinite Prandtl number limit. This infinite Prandtl number limit will remove the inertial term in the momentum equation \citep{hansen1990transition,constantin1999infinite}, filtering out shear instability, which can be seen by normalizing the velocity using the thermal diffusivity ($\boldsymbol{U}=\boldsymbol{U}^* H/\kappa_T$), resulting in the non-dimensional momentum equation as
\begin{equation}
    \frac{1}{Pr}\left(\frac{\partial\boldsymbol{U}}{\partial t} +\boldsymbol{U}\cdot\nabla\boldsymbol{U}\right) = -\nabla P + \nabla^2\boldsymbol{U} + Ra\left(T-\Lambda S\right)\hat{\mathbf{e}}_z.
    \label{eq.momentum_for_fin_Pr}
\end{equation}
At the infinite $Pr$ limit, there is no effect from the inertial terms ($\partial_t \boldsymbol{U}+\boldsymbol{U}\cdot\nabla\boldsymbol{U}$) in equation \eqref{eq.momentum_for_fin_Pr} \citep{doering2006bounds}, and thus, this limit filters out the possibility of shear instability. Thermohaline-shear instability in vertically periodic SDDC still occurs without the inertial term because this instability occurs due to the interaction between thermohaline convection and shear \citep{radko2016thermohaline}. However, the bifurcation diagram over the Prandtl number in Figure \ref{fig:tau_continuation}(b) shows that a similar thermohaline-shear instability \citep{radko2016thermohaline} does not occur in our vertically bounded domain with a linear background flow at the high Prandtl number limit.  

\begin{figure}
\centering
\includegraphics[width=1\linewidth]{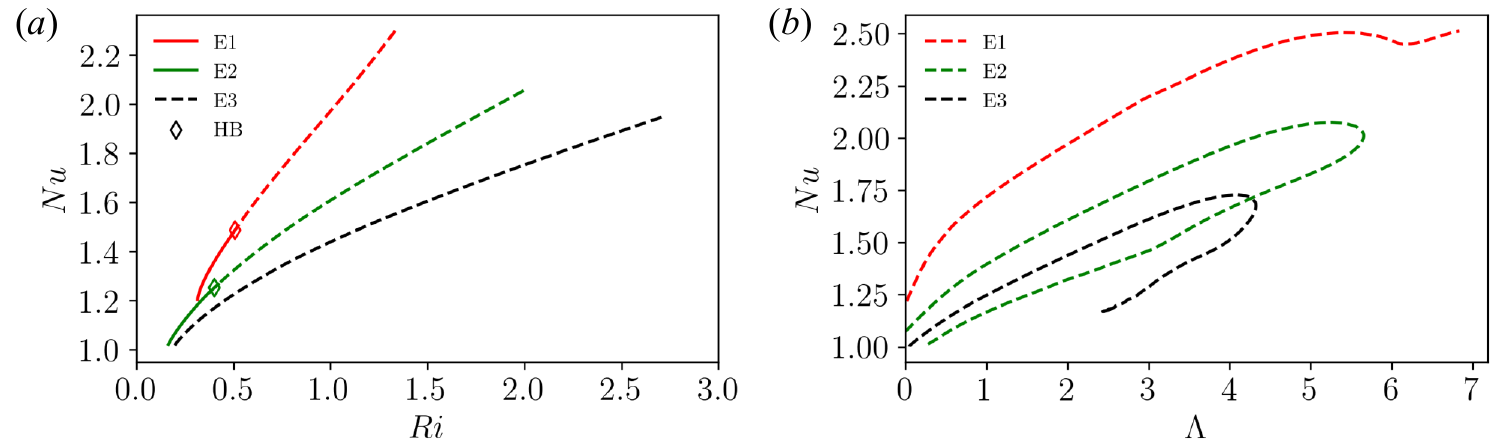}
\caption{Continuation of convective roll states E1-E3 over (a) the Richardson number $Ri$ associated with shear with fixed parameters $(Ra, \Lambda, Pr, \tau)=(10^5, 2, 7, 0.01)$ and (b) the density ratio $\Lambda$ with fixed parameters $(Ra, Ri, Pr, \tau)=(10^5, 1, 7, 0.01)$. The Nusselt number $Nu$ is used to visualize the continuations. The solid and dashed lines denote the stable and unstable solutions, respectively. The symbol $\Diamond$ indicates the Hopf bifurcation (HB) points.}	
\label{fig:Ri_Lambda_cont}
\end{figure}

To investigate the shear effects on the convective roll states, we conducted continuations of upper branch solutions of E1-E3 with respect to the change in the Richardson number $Ri=1/U_w^2$. Here, $U_w$ is the non-dimensional wall velocity difference. In figure \ref{fig:Ri_Lambda_cont}(a), we plot $Nu$ as functions of $Ri$. As shear strength increases (decreasing $Ri$), shear stabilizes convective roll states. E2 and E3 tend to approach the base state in which $Nu=Sh=1$ and $\gamma=\Lambda\tau$. The flow structure at the small $Ri$ regime is similar to the shear-influenced flow structures in the study by \citet{li2024double}. However, we cannot reach the strong shear limit ($Ri\rightarrow 0$) due to computational limits. When reducing the shear strength (increasing $Ri$), E1-E3 tend to reach convective states, leading to increased convective flux. The resulting fluxes for E1-E3 indicate a linearly increasing trend in the range of $Ri>1$. Here, our Nusselt number $Nu$ shows a monotonic behavior when changing $Ri$, while a non-monotonic dependence of $Nu$ on $Ri$ is observed in DNS for high $Ri$ \citep[Figure 8]{ravichandran2022combined} with the top wall as melting ice. In the study of \citet{li2024double}, the authors conducted numerous DNS and indicated that the laminar states are dominant under weak shear (high $Ri$) regimes. However, when we attempt to continue the $Ri$ continuation with higher $Ri$, the solutions of the convective roll states do not converge for all E1-E3 solutions. We will leave the numerical continuation over a wider $Ri$ regime as a direction for future work. 

The remaining dimensionless parameter governing the DDC system is the density ratio $\Lambda$. Figure \ref{fig:Ri_Lambda_cont}(b) shows $Nu$ as functions of $\Lambda$ for convective roll states E1-E3 on the upper branches in figure \ref{fig:eq_bifurcation} with parameters $(Ra, Ri, Pr, \tau)=(10^5, 1, 7, 0.01)$.  When $\Lambda=0$, the SDDC governed by equations \eqref{eq.sddc_fluctuation} becomes the sheared Rayleigh-B\'enard convection \citep{blass2020flow}. Figure \ref{fig:Ri_Lambda_cont}(b) shows that only E1 and E2 have $Nu>1$ when crossing $\Lambda=0$, indicating that they are connected to convective roll states in the sheared RBC. E3 solution instead connects to the conduction base state $(Nu=1)$ when $\Lambda=0$, which indicates that E3 exists only within SDDC. Moreover, figure \ref{fig:Ri_Lambda_cont}(b) shows fold bifurcations of E2-E3 solutions, while the $Nu$ number of E1 instead slightly drops around $\Lambda=6$ and then continues to increase again, approaching equilibrium states of higher flux. Here, E1 fails to converge for $\Lambda \geq 7$ in this study.

\section{Three-dimensional effects}\label{sec:3D_effect}
\begin{figure}
\centering
\includegraphics[width=1\linewidth]{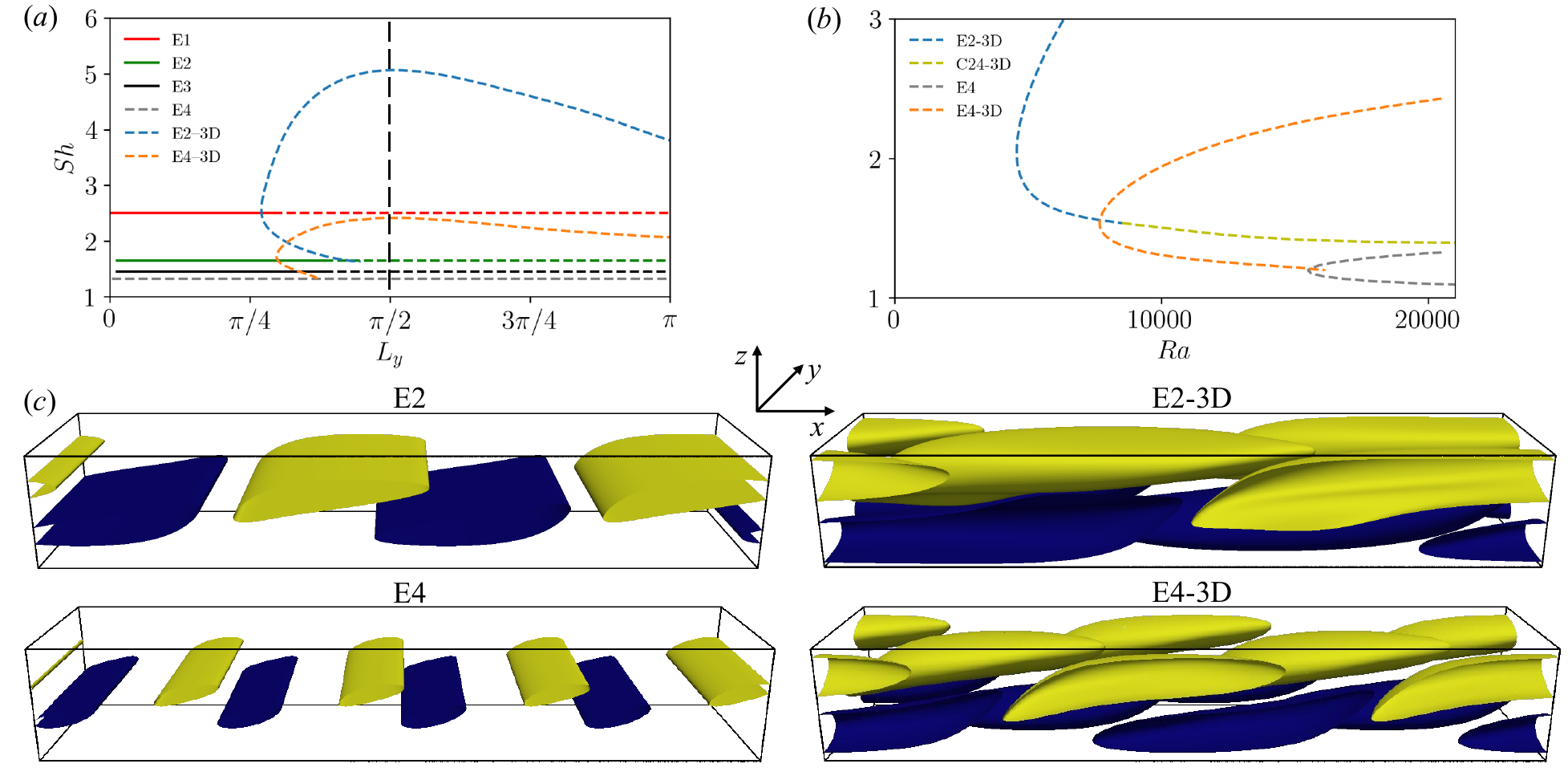}
\caption{Bifurcation diagram of three-dimensional convective roll states for fixed dimensionless parameters $(Ri, Pr, \tau, \Lambda)=(1, 7, 0.01, 2)$. (a) Bifurcation diagram in the change of spanwise domain size $L_y$ with $Ra=20000$. Three-dimensional convective roll states include E2-3D and E4-3D, where E4-3D bifurcated from transverse rolls corresponding to E4 via subcritical bifurcation. (b) Bifurcation diagram in the change of the Rayleigh number $Ra$ with spanwise domain size $L_y=\pi/2$. Solid lines indicate stable solutions, while dashed lines indicate unstable solutions. (c) Visualization of convective roll states in the 3D domain uses isosurfaces of temperature fluctuation $\theta$ with $L_y=\pi/2$ and $Ra=20000$. Here, yellow and blue colors indicate isosurfaces of $\theta=\pm0.1$ (E2 and E2-3D) and  $\theta=\pm0.08$ (E4 and E4-3D), respectively. Panels (b) and (c) use a fixed spanwise domain size of $L_y=\pi/2$.}	
\label{fig:3d_bifurcations}
\end{figure}

In this section, we investigate the three-dimensional (3D) effects on convective roll states for SDDC using a computational domain extended to the spanwise direction $L_y\in[0.005,\pi]$. The grid is fixed at $N_x\times N_y\times N_z=144\times64\times97$. We perform continuations in the change of $L_y$ for E1-E4, corresponding to equilibrium states with dimensionless parameters $(Ra, Ri, Pr, \tau, \Lambda)=(20000, 1, 7, 0.01, 2)$ on the upper branches in figure \ref{fig:eq_bifurcation}. At a minimal $L_y$, which can be considered a 2D domain, E1-E3 are stable while E4 is unstable, as shown in figure \ref{fig:3d_bifurcations}(a). As $L_y$ increases, 2D convective roll states will be unstable in the 3D domain with $L_y>0.9$ for E1 and $L_y>1.2$ for E2 and E3. Here, E4 is a fully unstable state at $Ra=20000$, even in the 2D domain. It shows that the reduction of spanwise length reduces spanwise instabilities, thereby stabilizing the transverse rolls.

At $Ra=20000$, E4 is unstable due to an oscillation instability represented by the leading eigenvalue as a complex conjugate pair. When $L_y$ increases, the first leading eigenvalue is replaced by an unstable real eigenvalue, leading to a subcritical bifurcation (see figure \ref{fig:3d_bifurcations}(a)). This bifurcation leads to a 3D convective roll state that satisfies the symmetry properties $\langle\pi_x,\pi_z,\tau_x(L_x/4),\tau_y(L_y/2)\rangle$, where $\lambda_y=L_y/2$ is an additional spanwise wavelength in equation \eqref{eq:sym_translation_y}. Because the 2D flow structure is broken, additional symmetry in the spanwise direction is necessary for 3D convective roll states, while this additional spanwise translation symmetry has no influence on the reflection symmetries ($\pi_x$ and $\pi_z$).  Here, we name the 3D solution satisfying $\langle\pi_x,\pi_z,\tau_x(L_x/4),\tau_y(L_y/2)\rangle$ symmetry as E4-3D, which corresponds to the E4 solution in 2D domain. This E4-3D solution is unstable in the entire subcritical bifurcation. In addition, another solution of the 3D convective roll state, E2-3D, has been found. We name it E2-3D because it satisfies the symmetry properties $\langle\pi_x,\pi_z,\tau_x(L_x/2),\tau_y(L_y/2)\rangle$. Both E2-3D and E4-3D reach maximum fluxes close to $L_y=\pi/2$. These 3D solutions generate streamwise elongated structures near the wall, which resemble scallop structures in field observations within ice-shelf cavities \citep[Figure 4]{washam2023direct}. Although the continuation of E2-3D exhibits subcritical behavior, it does not connect directly to the transverse rolls of E2 in figure \ref{fig:3d_bifurcations}(a). Additionally, we have conducted continuations in the change of $Ra$ with a fixed domain of $L_x\times L_y\times L_z=2\pi\times \pi/2\times 1$. As shown in figure \ref{fig:3d_bifurcations}(b), E2-3D becomes a three-dimensional mixed mode (C24-3D) on its lower branch, which is similar to the way in which E2 becomes C24 discussed in section \ref{sec:mixed_modes}. This mixed mode C24-3D brings the same qualitative flow behavior as E4-3D. Although E4-3D bifurcates from E4 as $L_y$ extends, it does not occur during the $Ra$ continuation at these selected parameters. To illustrate the three-dimensional flow structure of the convective roll states in the 3D domain, we then show isosurfaces of temperature fluctuations for E2, E4, E2-3D, and E4-3D corresponding to $Ra=20000$ and $L_y=\pi/2$ in figure \ref{fig:3d_bifurcations}(c). The 3D convective roll states have elongated flow structures in the streamwise direction, which are qualitatively similar to the longitudinal rolls \citep{clever1977instabilities_PCF,clever1977instabilities_ILC,reetz2020invariant}. 

\begin{figure}
\centering
\includegraphics[width=1\linewidth]{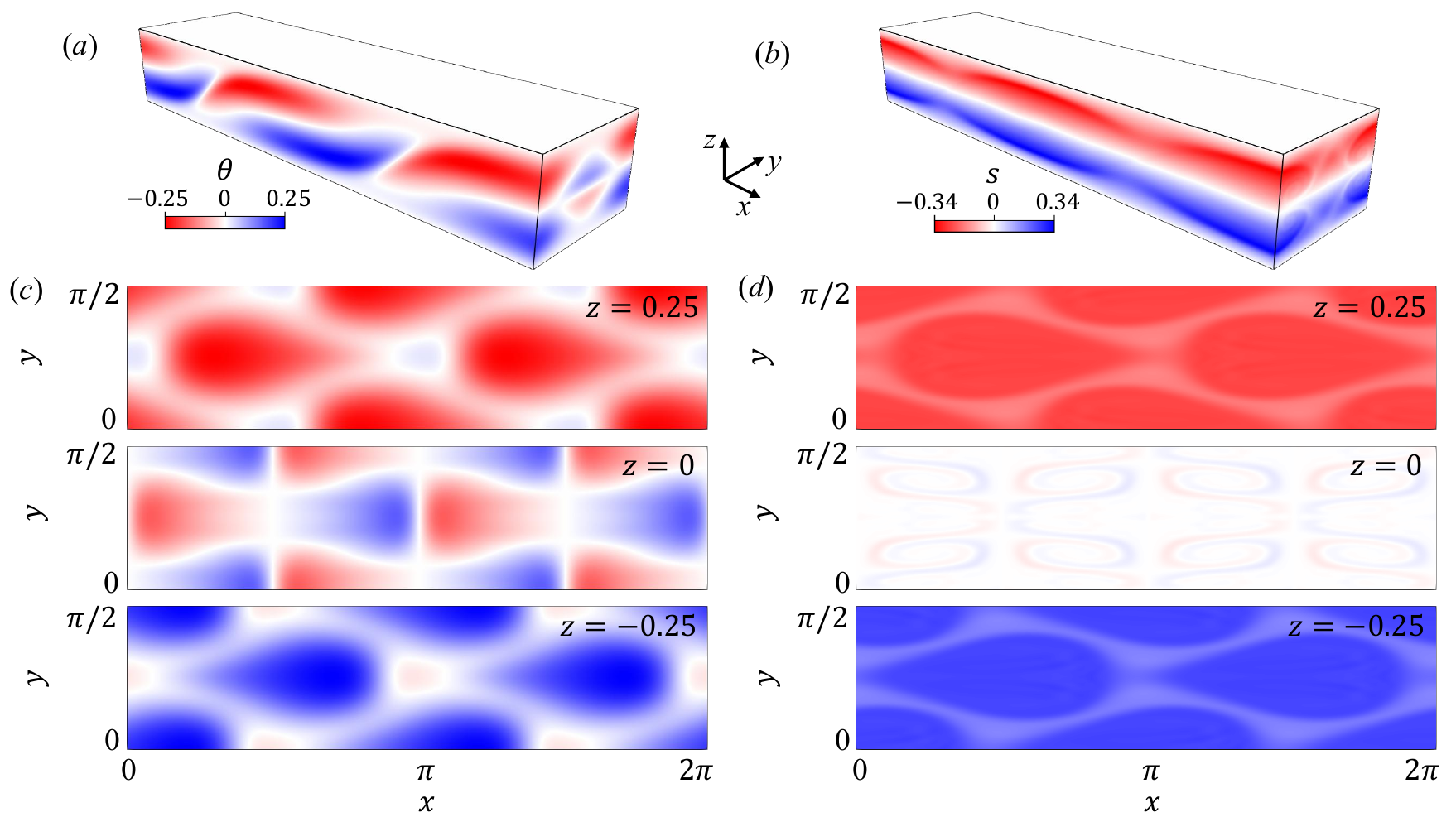}
\caption{Key properties of E4-3D displaying 3D view of (a) temperature fluctuations $\theta$ and (b) salinity fluctuations $s$. Panels (c)-(d) show three different slices of horizontal plane ($x,y$) at $z=0.25, 0$, and $-0.25$ for (c) temperature fluctuations $\theta$ and  (d) salinity fluctuations $s$. The dimensionless parameters for this solution are $(Ra, Ri, Pr, \tau, \Lambda)=(2\times 10^4,1, 7, 0.01, 2)$. The color ranges for the slices and the 3D view panels are identical, respectively. The solution is obtained in a fixed domain size of $L_y=\pi/2$.}	
\label{fig:3D_convection_rolls}
\end{figure}

\begin{figure}
\centering
\includegraphics[width=1\linewidth]{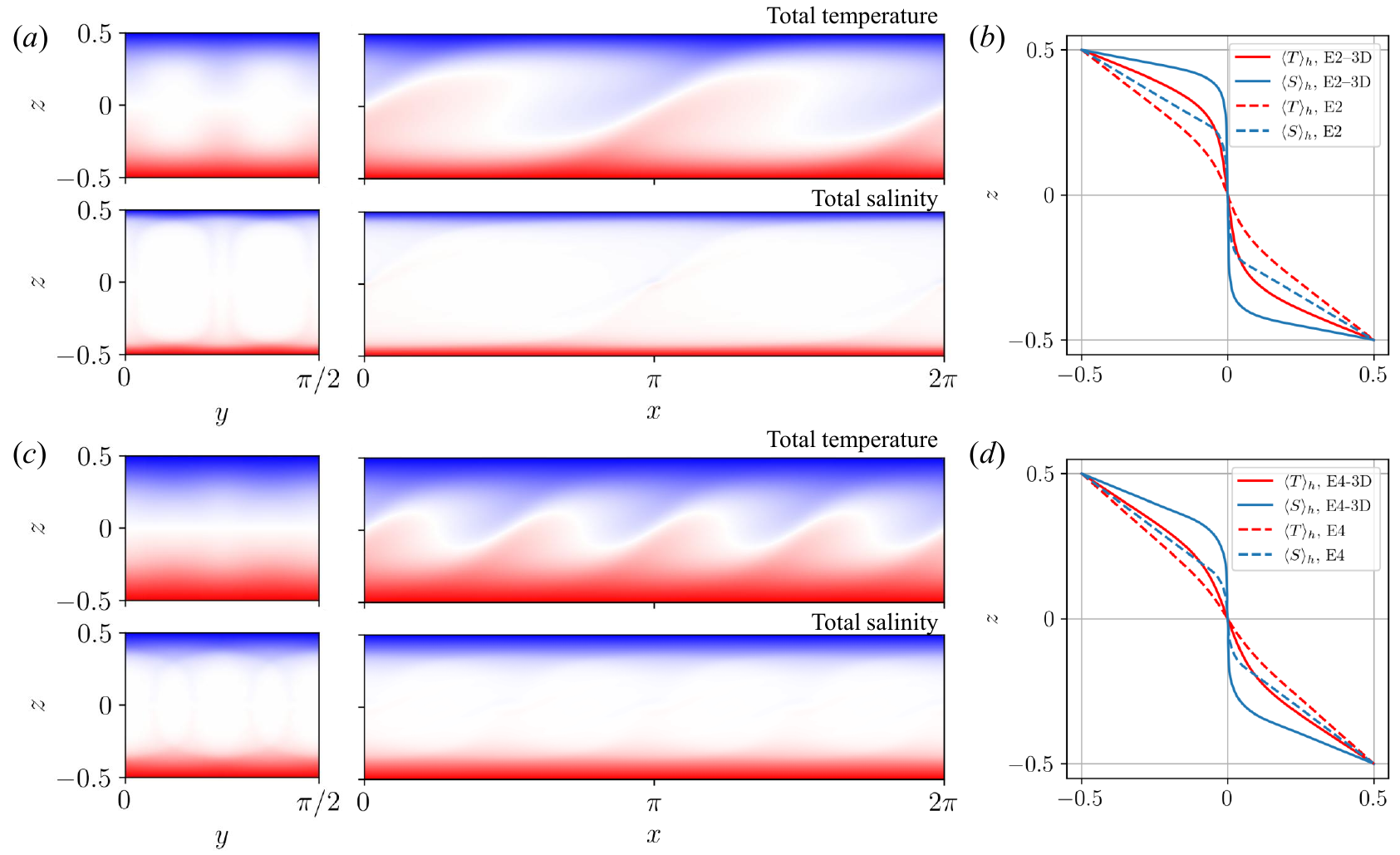}
\caption{Averaged profiles of (a) E2-3D and (c) E4-3D for SDDC with dimensionless parameters $(Ra, Ri, Pr, \tau, \Lambda)=(20000, 1, 7, 0.01, 2)$ and domain size are $L_x\times L_y\times L_z=2\pi\times \pi/2\times 1$. (a,c) Streamwise-averaged (left) and spanwise-averaged (right) total temperature $T$ (top) and total salinity $S$ (bottom). Data is plotted in a scalar value range of $[-0.5,0.5]$ from blue to red color. Panels (b) and (d) compare horizontally-averaged scalar profiles between 2D and 3D convective roll states.}	
\label{fig:3D_mean_profiles}
\end{figure}

Figure \ref{fig:3D_convection_rolls} shows the 3D views (panels (a,b)) and the horizontal slices (panels (c,d)) of temperature and salinity fluctuations of E4-3D for $L_y=\pi/2$ and $(Ra, Ri, Pr, \tau, \Lambda)=(2\times 10^4,1, 7, 0.01, 2)$. At $z=\pm0.25$, panels (c)-(d) of figure \ref{fig:3D_convection_rolls} show that the distributions of both temperature and salinity exhibit staggered droplet-like structures inclined in the streamwise direction. At the mid-plane $z=0$, the scalar fluctuations exhibit a symmetric distribution. Figure \ref{fig:3D_mean_profiles} shows the properties of the averaged profiles of temperature and salinity for E2-3D and E4-3D. Streamwise-averaged (left) and spanwise-averaged (right) scalar distributions are plotted in figure \ref{fig:3D_mean_profiles}(a,c), respectively. The spanwise-averaged distributions confirm their translation symmetries. Also, E2-3D has well-mixed regions larger than those of E4-3D in both streamwise-averaged and spanwise-averaged profiles. Enhancement of mixing in a 3D domain is emphasized by the larger well-mixed region of mean profiles of E2-3D and E4-3D compared with their 2D counterparts, E2 and E4, as shown in figure \ref{fig:3D_mean_profiles}(b,d). 

The present results of convection rolls in a 2D domain correspond to transverse rolls, where the roll axes are orthogonal to the imposed shear flow. Shear can stabilize these transverse rolls but has no or minimal effect on longitudinal rolls, where the roll axes are parallel to the mean flow direction \citep{clever1977instabilities_PCF,clever1977instabilities_ILC,caltagirone1985solutions,wen2018inclined,reetz2020invariant}. As a result, oscillatory instability of the diffusive regime DDC \citep{radko2013double} in the form of longitudinal rolls is still possible in sheared DDC. Consequently, more complex flow dynamics are anticipated in three dimensions, particularly in the form of longitudinal/streamwise rolls resembling those observed in PCF \citep{kawahara2001periodic,halcrow2008charting,gibson2009equilibrium,engel2025search} and the inclined Rayleigh-B{\'e}nard convection \citep{reetz2019exact,reetz2020invariant,reetz2020invariant_part2}.

\section{Conclusions}\label{sec:conclusion}

This work computed exact coherent structures (ECS) of sheared double-diffusive convection (SDDC) in the diffusive regime (cold fresh water on top of hot salty water) with a uniform background shear. We chose dimensionless parameters relevant to the oceanographic settings, except when conducting continuations of governing parameters. In most cases, we have the Prandtl number $Pr=7$, the diffusivity ratio $\tau=0.01$, the density ratio $\Lambda=2$, and the Richardson number $Ri=1$.

DNS results show that the layering behavior in SDDC is robust in both the no-slip boundary condition analyzed here and the stress-free boundary condition \citep{yang2022layering,li2024double}. This layering mechanism leads to dominant one-layer convection as the final state. A decrease in the convective fluxes is observed in the no-slip condition compared with the stress-free boundary condition. Our results show that SDDC has a transition from laminar conductive flow to steady convective states and eventually to unsteady states. Steady convective states have one-layer roll structures, and the instantaneous one-layer roll structures are still captured until $Ra\lessapprox 2\times 10^5$ for unsteady states. The transition from steady to unsteady flow leads to non-monotonic behavior in $Nu$. The flux ratio at $Ra\lessapprox 2\times 10^5$ is smaller than $\sqrt{\tau}$ associated with the DDC-dominated regime \citep{linden1978diffusive}, but larger than the value $\gamma=\Lambda\tau$ associated with the conductive base state, which indicates the shear-influenced regime \citep{li2024double}. 

In this study, the 2D ECS of the equilibrium solutions and periodic orbits are identified within the shear-influenced regime. We found the co-existence of steady convective-roll states E$n$ ($n=1, 2, 3$), representing flow structures of $n$ identical counter-rotating convection roll pairs. The convective fluxes and area of well-mixed regions decrease with the increase in the number of convective rolls. The flux ratio obtained from multiple-roll states E1-E3 is much smaller than the $\sqrt{\tau}$ associated with the DDC-dominated regime \citep{linden1978diffusive}, emphasizing the properties of the shear-influenced regime \citep{li2024double}. Under the continuation of $Nu$ as a function of $Ra$, multiple-roll states undergo saddle-node bifurcations, with a stable upper branch and an unstable lower branch near the saddle-node bifurcations. The $Ra_{\text{SN}}$ associated with the saddle-node bifurcation follows the trend of $Ra_{\text{SN}}(E3)<Ra_{\text{SN}}(E1)<Ra_{\text{SN}}(E2)$. The lower branch of E3 approaches the base state, suggesting a subcritical bifurcation from the base state, while the lower branches of E1 and E2 connect to the corresponding mixed modes (C12 and C24). Here,  mixed modes represent flow structures that are a combination of small and large rolls, connecting multiple-roll states associated with different pairs of counter-rotating rolls. For example, the mixed-mode C12 connects the lower branch of E1 state to the upper branch of E2 convective-roll states. 

Hopf bifurcation occurs on the stable upper branches leading to periodic orbits. Here, we identified periodic orbits PO1 and PO2, corresponding to the Hopf bifurcation on the upper branches of E1 and E2 solutions as increasing $Ra$, respectively. They have similar oscillation characteristics in the form of standing waves. For these periodic orbits, the instantaneous value of $Sh$ is significantly higher than that of $Nu$, even across negative values. Unlike the equilibrium solutions, the instantaneous flux ratio of periodic orbits surpasses the value of $\gamma=\sqrt{\tau}$ \citep{linden1978diffusive}, indicating that periodic orbits have certain moments resembling DDC-dominated states \citep{li2024double}. For higher $Ra$, the DNS of the SDDC exhibits chaotic behavior but remains confined within invariant manifolds associated with the identified equilibrium solutions and periodic orbits.

In parameter continuation of $\tau$ and $Pr$, the 2D convective-roll states exhibit subcritical behavior, including the saddle-node bifurcation and the Hopf bifurcation on the upper branch of steady solutions. The enhancement of convective fluxes increases with a decrease in the number of convection rolls. Under the change of shear strength, it shows that strong shear (small $Ri$) stabilizes these convective roll states and reduces convective fluxes. E1 and E2 connect directly to steady-state solutions of the wall-sheared Rayleigh-B\'enard convection, as demonstrated by the continuation of the density ratio. We also performed continuations of the spanwise domain size to investigate the 3D behavior, where the transverse rolls (convective-roll states) that are stable within a 2D domain can become unstable as the spanwise domain size increases. The 3D multiple-roll states E2-3D and E4-3D exhibit subcritical behavior. In particular, the E4-3D state is generated from the subcritical bifurcation of the corresponding 2D transverse roll state E4.

The present study has focused on the subcritical transitional regime of SDDC in the form of one-layer convection rolls. The route to chaotic behavior \citep{yang2022layering,li2024double} may be governed by a combination of many unstable ECS. In particular, finding unstable periodic orbits can be facilitated by other advanced techniques, such as the variational methods \citep{parker2022variational,azimi2022constructing} and the multi-shooting method \citep{sanchez2010multiple,van2011matrix}. An unstable periodic orbit predictor based on dynamic mode decomposition has also been shown to generate excellent guesses for finding periodic orbits \citep{engel2025search}. Moreover, finding ghost states using the variational method will also be interesting to uncover the underlying dynamics at parameter regimes where ECS no longer exists \citep{zheng2025ghost}.

Moreover, the bifurcation diagram presented here does not explain the dominant flow structures beyond $Ra\gtrapprox 2\times 10^5$. It is necessary to promote further research to obtain ECS for such flow structures at higher Rayleigh number regimes. In addition, the ECS of multiple layers, suggested by DNS \citep{yang2022layering,li2024double}, is also interesting to explore. However, this case often generates multiple sharp interfaces with large gradients, leading to challenges in finding ECS in the form of multiple layers. The SDDC in the salt-finger regime is also of our interest \citep{linden1974salt,paparella1999sheared,fernandes2010salt,li2022flow}, and the extension to the vertical periodic domain in both the diffusive and salt-finger regimes will be studied in future research.

\backsection[Acknowledgements]{We thank Zheng Zheng and Jake Langham for helpful discussions regarding the use of Channelflow 2.0 code. The computational work for this project was conducted using resources provided by the Storrs High-Performance Computing (HPC) cluster. We extend our gratitude to the UConn Storrs HPC and its team for their resources and support, which aided in achieving these results. The computational resources for this project were also provided by the NSF ACCESS program (project number: PHY240243), allowing us to use the Bridges-2 cluster in the Pittsburgh Supercomputing Center.}

\backsection[Funding]{This research was supported by the University of Connecticut (UConn) Research Excellence Program (REP).}

\backsection[Declaration of interests]{The authors report no conflict of interest.}

\backsection[Data availability statement]{The data and post-processing code that support the findings of this study are openly available in Zenodo at \textcolor{blue}{https://doi.org/10.5281/zenodo.17956971}. The ChFlow-DDC code is available at \textcolor{blue}{https://doi.org/10.5281/zenodo.17956880}.}

\backsection[Author ORCIDs]{\\
Van Duc Nguyen: \href{https://orcid.org/0000-0001-5864-5555}{https://orcid.org/0000-0001-5864-5555}\\
Chang Liu: \href{https://orcid.org/0000-0003-2091-6545}{https://orcid.org/0000-0003-2091-6545}}


\appendix

\section{Validations of ChFlow-DDC code}\label{app:validations}

\begin{figure}
    \centering
    \includegraphics[width=1\linewidth]{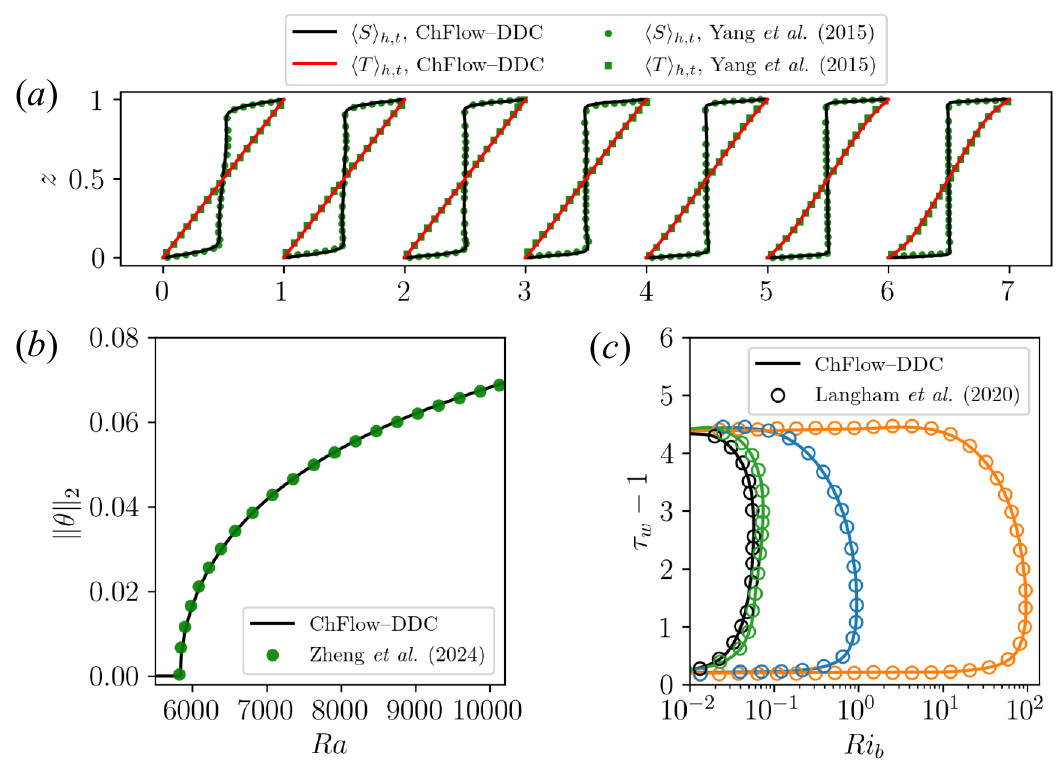}
    \caption{Validations of ChFlow--DDC code: (\textit{a}) Mean profiles of total temperature $T$ and salinity $S$ for DNS of DDC in salt-finger regime with parameters $Ra=10^6$, $Pr=7$, $\tau=0.01$, and various density ratio $R_\rho=1/\Lambda=10, 5, 2, 1, 0.5, 0.2$, and 0.1 from the left to the right. Red and black lines represent temperature and salinity obtained from ChFlow-DDC, whereas the green circle $\circ$ and square $\square$ indicate the corresponding results of \cite[Figure 2b]{yang2015salinity}. Results of each case are shifted rightward by 1 from the previous left one. (\textit{b}) Supercritical pitchfork bifurcation of four-roll equilibrium solution named FP1 for vertical convection \citep[Figure 2a]{zheng2024natural}. Bifurcation diagram is plotted by $L_2$-norm of temperature fluctuation $\|\theta\|_2$ defined by equation \eqref{eq.l2norm_theta}. (\textit{c}) Parameter continuation of the $EQ_7$ (lower) and $EQ_8$ (upper)  equilibrium states for the stratified PCF \citep[Figure 2a]{langham2020stably} using wall-shear stress ($\tau_w-1$, where $\tau_w$ is defined by equation \eqref{eq:wallshear_stress}) in the change of bulk Richardson number for different Prandtl numbers: $Pr=10^{-4}$ (orange), $Pr=10^{-2}$  (blue), $Pr=1$ (black), and $Pr=7$ (green).}
    \label{fig:validation}
\end{figure}

To validate the code, we implement both DNS and parameter continuation of equilibrium solutions for several flow problems. First of all, we perform seven 3-D DNS of DDC in finger mode corresponding to the cases of $Ra=10^6$ in the work of \cite[Figure 2b]{yang2015salinity}, as shown in figure \ref{fig:validation}(a). The horizontally averaged temperature and salinity profiles obtained from ChFlow-DDC are consistent with DNS results from \citet[Figure 2b]{yang2015salinity}. In particular, both the DNS results from ChFlow-DDC and \citet{yang2015salinity} show that there is a well-mixed region of mean total salinity profile in the middle of the domain. The mean total temperatures are close to a linear profile and display a visible deviation from the linear profile at lower density ratios $R_\rho:=1/\Lambda$ ($R_\rho=0.2$ and 0.1). This demonstrates the ability of ChFlow-DDC to perform DNS for DDC problems. 

Second, we reproduce the supercritical pitchfork bifurcation of the four-roll state named FP1 in natural convection \citep[Figure 2a]{zheng2024natural}. The $L_2$-norm of temperature fluctuation
\begin{equation}
    \|\theta\|_2=\left(\frac{1}{L_x L_y L_z} \iiint\theta^2 dxdydz \right)^{1/2}
    \label{eq.l2norm_theta}
\end{equation}
is used to validate ChFlow-DDC with results in \citep{zheng2024natural}. As shown in figure \ref{fig:validation}(b), the result shows excellent agreement along the bifurcation curve with an increase of $Ra$. Our code indicates that the bifurcation point is $Ra=5839$, which is close to $Ra=5826$ by \cite{zheng2024natural} and the threshold 5800 using numerical simulation measurements by \cite{gao2013transition}. This demonstrates the ability of the code to determine the equilibrium state and parameter continuation. 

Finally, we verify the control of different parameters for the governing equations of the system by using the continuation between two equilibrium states (EQ7 and EQ8) for the stratified PCF \citep[Figure 2a]{langham2020stably}. Here, four Prandtl numbers $10^{-4}$, $10^{-2}$, $1$, and 7 are investigated. Our results are overlapping with those of the reference, see figure \ref{fig:validation}(c). All of these validations demonstrate that the ChFlow-DDC code is accurate, effective, and suitable for the present study.

\section{Mesh independence studies}\label{app:mesh_ind_study}
We have conducted two mesh independence studies to demonstrate that our selected mesh is suitable for this study. First, we performed a series of DNS with varying grid resolutions in Table \ref{tab:mesh_ind_dns}. Computations are performed for parameters $(Ra, Ri, Pr, \tau, \Lambda)=(10^5, 1, 7, 0.01, 2)$, where convection rolls appear at sufficiently high $Ra$. The table shows the grid convergence as the resolution increases. Also, we compare these results obtained from Chflow-DDC with pure 2D DNS using Dedalus v3 \citep{burns2020dedalus}. We select a mesh of $128\times97$, which is efficient and minimal in cost but shows only a difference of 1\% compared to the largest mesh. In general, this parameter set represents the chaotic flow regime discussed in section \ref{sec:dns.2}. Time-averaged statistics from DNS may not show clear convergence but still remain bounded by an acceptable difference of 5\%. In Table \ref{tab:mesh_ind_nsolver}, a similar test for grid convergence of equilibrium solutions E1-E3 is performed. The $128\times97$ mesh exhibits a deviation of under 1\% relative to the largest mesh. These tests demonstrate the ability of our selected mesh for both DNS and computing ECS.

\begin{table}
  \begin{center}
\def~{\hphantom{0}}
  \begin{tabular}{lccccc}
      Code & $N_x$ & $N_z$ & $\overline{Nu}$ & $\overline{Sh}$ & $\overline{\gamma}$ \\[3pt]
       ChFlow-DDC & 96 & 65 & 1.734 (0.1\%) & 3.591 (1.5\%) & 0.041 \\
       ChFlow-DDC & 96 & 81 & 1.766 (1.7\%) & 3.806 (4.4\%) & 0.043 \\
       ChFlow-DDC & 128 & 97 & 1.752 (0.9\%, 0.6\%) & 3.619 (0.7\%, 1\%) & 0.041 \\
       ChFlow-DDC & 144 & 97 & 1.720 (0.9\%) & 3.587 (1.5\%) & 0.042 \\
       ChFlow-DDC & 144 & 113 & 1.776 (2.3\%) & 3.764 (3.2\%) & 0.042 \\
       ChFlow-DDC & 192 & 129 & 1.709 (1.5\%) & 3.698 (1.4\%) & 0.043 \\
       ChFlow-DDC & 256 & 129 & 1.729 (0.4\%, 2.8\%) & 3.665 (0.5\%, 0.8\%) & 0.042 \\
       ChFlow-DDC & 256 & 161 & 1.736 (0\%, 0.8\%) & 3.645 (0\%, 0.7\%) & 0.042 \\
       Dedalus v3 & 128 & 97 & 1.740 & 3.658 & 0.042 \\
       Dedalus v3 & 256 & 129 & 1.759 & 3.695 & 0.042 \\
       Dedalus v3 & 256 & 161 & 1.722 & 3.673 &  0.042 \\
  \end{tabular}
  \caption{Mesh independence study using DNS for fixed parameters $(Ra, Ri, Pr, \tau, \Lambda)=(10^5, 1, 7, 0.01, 2)$ in a domain of $L_x\times L_y\times L_z=2\pi\times 0.005\times 1$. $N_x$ and $N_z$ indicate the resolution in the streamwise ($x$) and vertical ($z$) directions, respectively. Spanwise resolution $N_y$ is fixed at 10 grid points. Time-averaged results obtained from ChFlow-DDC are compared to those from pure 2D SDDC using Dedalus v3 \citep{burns2020dedalus}. The other three columns show the time average of the Nusselt number ($Nu$), the Sherwood number ($Sh$), and the flux ratio ($\gamma$). The first percentage number in parentheses indicates the difference compared to the results of the largest mesh using ChFlow-DDC, while the second percentage value in parentheses indicates the relative difference comparing the results obtained from 2D DDC by ChFlow-DDC against those obtained by Dedalus with the corresponding mesh. }
  \label{tab:mesh_ind_dns}
  \end{center}
\end{table}

\begin{table}
  \begin{center}
\def~{\hphantom{0}}
  \begin{tabular}{cccccccc}
      &&\multicolumn{2}{c}{E1} & \multicolumn{2}{c}{E2} & \multicolumn{2}{c}{E3} \\[3pt]
      $N_x$ & $N_z$ & $Nu$ & $Sh$ & $Nu$ & $Sh$ & $Nu$ & $Sh$\\
       96 & 65 & 1.981 (0.6\%) & 2.738 (0.6\%) & 1.610 (0.2\%) & 1.974 (1.1\%) & 1.436 (0.1\%) & 1.602 (4\%) \\
       128 & 97 & 1.971 (0.1\%) & 2.722 (0\%) & 1.607 (0.1\%) & 1.951 (0.1\%) & 1.439 (0.3\%) & 1.680 (0.7\%) \\
       192 & 129& 1.970 (0\%) & 2.723 (0.1\%) & 1.606 (0\%) & 1.953 (0.1\%) & 1.435 (0\%) & 1.662 (0.4\%) \\
       256 & 129& 1.970 (0\%) & 2.724 (0.1\%) & 1.606 (0\%) & 1.951 (0.1\%) & 1.435 (0\%) & 1.667 (0.1\%) \\
       256 & 161 & 1.970 & 2.721 & 1.606 & 1.952 & 1.435 & 1.668 \\
  \end{tabular}
  \caption{Mesh independence study for grid convergence of equilibrium solutions of 2D convective roll states E1-E3 at fixed parameters $(Ra, Ri, Pr, \tau, \Lambda)=(10^5, 1, 7, 0.01, 2)$. The percentage number in parentheses indicates the difference compared to the results of the largest mesh.}
  \label{tab:mesh_ind_nsolver}
  \end{center}
\end{table}

\section{Statistical data of direct numerical simulations}\label{app:dns_data}
Table \ref{tab:dns_data} lists the statistical results obtained from DNS for SDDC at fixed parameters $(Ri, Pr, \tau, \Lambda)=(1, 7, 0.01, 2)$, which are discussed in figure \ref{fig:dns_flow_pattern}. The $Ra$ varies within a range of $[10^4,10^6]$. We use a sine-function condition ($-\delta \sin{(2\pi k_z z)}$) for the initial fluctuations of temperature and salinity, where $\delta$ and $k_z$ represent the amplitude and vertical wavenumber. We fix $k_z=1$ to get one-layer convection immediately. Readers may refer to the work of \citet{li2024double} to get a more detailed setup.
\begin{table}
  \begin{center}
\def~{\hphantom{0}}
  \begin{tabular}{lcccccccc}
      $Ra$ & $N_x$ & $N_z$ & $\delta$ & $\overline{Nu}$ & $\overline{Sh}$ & $\overline{\gamma}$ & $t_{stat}$ \\[3pt]
       $10^4$   & 128 & 97 & 0.25 & 1 & 1 & 0.2 & 5000 \\
       $1.5\times10^4$   & 128 & 97 & 0.25 & 1.075 & 1.283 & 0.024 & 5000 \\
       $2\times10^4$   & 128 & 97 & 0.25 & 1.280 & 1.654 & 0.026 & 3000\\
       $3\times10^4$   & 128 & 97 & 0.25 & 1.821 & 2.797 & 0.031 & 3000 \\
       $4\times10^4$  & 128 & 97 & 0.25 & 1.512 & 2.640 & 0.035 & 8000 \\
       $5\times10^4$  & 128 & 97 & 0.25 & 1.550 & 2.864 & 0.037 & 2500 \\
       $10^5$  & 256 & 129 & 0.15 & 1.701 & 3.744 & 0.044 & 4000 \\
       $1.5\times10^5$  & 256 & 129 & 0.15 & 1.977 & 3.671 & 0.037 & 4000 \\
       $2\times10^5$  & 256 & 129 & 0.15 & 2.221 & 3.751 & 0.034 & 4000 \\
       $3\times10^5$  & 256 & 129 & 0.15 & 2.517 & 4.428 & 0.035 & 4000 \\
       $5\times10^5$  & 256 & 129 & 0.15 & 2.846 & 5.535 & 0.039 & 4000\\
       $10^6$  & 256 & 129 & 0.15 & 3.480 & 7.782 & 0.045 & 4000 
  \end{tabular}
  \caption{Key results obtained from statistics of DNS for SDDC in the rise of $Ra$ with fixed parameters $(Ri, Pr, \tau, \Lambda)=(1, 7, 0.01, 2)$. $t_{stat}$ indicates time intervals of average.}
  \label{tab:dns_data}
  \end{center}
\end{table}

\section{Linear stability analysis of base solution for SDDC}\label{app:LSA_base_SDDC}
This section performs the linear stability analysis of the base state \eqref{eq.base_state} in a wall-bounded SDDC, which supports the results discussed in figures \ref{fig:eq_bifurcation} and \ref{fig:EQPO_bifurcation_diagram}. To examine the linear stability of the base state, the governing equations \eqref{eq.sddc_fluctuation} are linearized with respect to the base state \eqref{eq.base_state} as expressed by
\begin{subequations}
\begin{align}
\boldsymbol{\nabla}\cdot \boldsymbol{u} &= 0,\\
\frac{\partial u}{\partial t} &= -U_b\frac{\partial u}{\partial x} - w\frac{d U_b}{d z} -\frac{\partial p}{\partial x} + \sqrt{\frac{Pr}{Ra}}\nabla^2 u,\\
\frac{\partial v}{\partial t} &= -U_b\frac{\partial v}{\partial x} -\frac{\partial p}{\partial y} + \sqrt{\frac{Pr}{Ra}}\nabla^2 v,\\
\frac{\partial w}{\partial t} &= -U_b\frac{\partial w}{\partial x} -\frac{\partial p}{\partial z} + \sqrt{\frac{Pr}{Ra}}\nabla^2 w +(\theta-\Lambda s),\\
\frac{\partial \theta}{\partial t} &= -U_b\frac{\partial \theta}{\partial x} + w + \frac{1}{\sqrt{PrRa}}\nabla^2 \theta,\\
\frac{\partial s}{\partial t} &= -U_b\frac{\partial s}{\partial x} + w + \frac{\tau}{\sqrt{PrRa}}\nabla^2 s,
\end{align}
\end{subequations}
where velocity, temperature, salinity, and pressure fluctuations are defined in equation \eqref{eq:base_state}. Here, $U_b=Ri^{-1/2}z$ and $U_b^\prime=\frac{d U_b}{d z}=Ri^{-1/2}$ are the base flow and its derivative over $z$. The generalized eigenvalue problem is expressed in matrix form
\begin{equation}
    \boldsymbol{B}\frac{\partial \boldsymbol{\xi}}{\partial t}=\boldsymbol{A}\boldsymbol{\xi}
\end{equation}
with 
\begin{subequations}
\begin{align}
    \boldsymbol{\xi}&=[u,v,w,p,\theta,s]^T,\\
    \boldsymbol{A} &= \begin{bmatrix}
\boldsymbol{M}_u & \boldsymbol{0} & -U_b^\prime & -\partial_x & \boldsymbol{0} & \boldsymbol{0}\\
\boldsymbol{0} & \boldsymbol{M}_u & \boldsymbol{0} & -\partial_y  & \boldsymbol{0} & \boldsymbol{0}\\
\boldsymbol{0} & \boldsymbol{0} & \boldsymbol{M}_u & -\partial_z & \boldsymbol{I} & -\Lambda \\
\partial_x  & \partial_y  & \partial_z & \boldsymbol{0} & \boldsymbol{0} & \boldsymbol{0}\\
\boldsymbol{0} & \boldsymbol{0} & \boldsymbol{I} & \boldsymbol{0} & \boldsymbol{M}_\theta & \boldsymbol{0}\\
\boldsymbol{0} & \boldsymbol{0} & \boldsymbol{I} & \boldsymbol{0} & \boldsymbol{0} & \boldsymbol{M}_s
 \end{bmatrix}, \label{eq:matrix_A} \\
 \boldsymbol{B} &= \begin{bmatrix}
    \boldsymbol{I} & \boldsymbol{0} & \boldsymbol{0} & \boldsymbol{0} & \boldsymbol{0} & \boldsymbol{0}\\
    \boldsymbol{0} & \boldsymbol{I} & \boldsymbol{0} & \boldsymbol{0} & \boldsymbol{0} & \boldsymbol{0}\\
    \boldsymbol{0} & \boldsymbol{0} & \boldsymbol{I} & \boldsymbol{0} & \boldsymbol{0} & \boldsymbol{0}\\
    \boldsymbol{0} & \boldsymbol{0} & \boldsymbol{0} & \boldsymbol{0} & \boldsymbol{0} & \boldsymbol{0}\\
    \boldsymbol{0} & \boldsymbol{0} & \boldsymbol{0} & \boldsymbol{0} & \boldsymbol{I} & \boldsymbol{0}\\
    \boldsymbol{0} & \boldsymbol{0} & \boldsymbol{0} & \boldsymbol{0} & \boldsymbol{0} & \boldsymbol{I}
    \end{bmatrix},
\end{align}
\end{subequations}
where $\boldsymbol{0}$ and $\boldsymbol{I}$ are the zero and identity matrices. The other terms in the matrix $\boldsymbol{A}$ of equation \eqref{eq:matrix_A} are defined as follows:
\begin{subequations}
\begin{align}
    \boldsymbol{M}_u &=-U_b\partial_x+\sqrt{\frac{Pr}{Ra}}\nabla^2, \\
    \boldsymbol{M}_\theta &=-U_b\partial_x+\frac{1}{\sqrt{PrRa}}\nabla^2, \\
    \boldsymbol{M}_s &=-U_b\partial_x+\frac{\tau}{\sqrt{PrRa}}\nabla^2. 
\end{align}
\end{subequations}

We then use the normal mode ansatz to assume fluctuations of velocity, temperature, salinity, and pressure as harmonic functions in the horizontal direction associated with a wavenumber pair ($k_x$, $k_y$) and the complex amplitudes $\hat{(\cdot)}$ as follows:
\begin{subequations}
    \begin{align}
        \boldsymbol{u}(x, y, z, t) &= \hat{\boldsymbol{u}}(z, t) e^{\text{i}(k_x x + k_y y)} + \text{c.c.}, \\
        \theta(x, y, z, t) &= \hat{\theta}(z, t) e^{\text{i}(k_x x + k_y y)} + \text{c.c.}, \\
        s(x, y, z, t) &= \hat{s}(z, t) e^{\text{i}(k_x x + k_y y)} + \text{c.c.}, \\
        p(x, y, z, t) &= \hat{p}(z, t) e^{\text{i}(k_x x + k_y y)} + \text{c.c.},
    \end{align}
\end{subequations}
where c.c. indicates a complex conjugate. These definitions lead to $\hat{\partial}_x=\text{i}k_x$, $\hat{\partial}_y=\text{i}k_y$, and $\hat{\nabla}^2:=\partial_z^2-(k_x^2+k_y^2)$. A Chebyshev collocation method is used in the $z$-direction with $\partial_z$ represented by the Chebyshev differentiation matrix \citep{weideman2000matlab}. The generalized eigenvalue problem is solved by the function $\mathrm{eig(\boldsymbol{A},\boldsymbol{B})}$ in \textsc{MATLAB}. First, we compute the linear stability for SDDC with fixed parameters $(Ra, Ri, Pr, \tau, \Lambda)=(10^7, 1, 7, 0.01, 2)$. The number of Chebyshev collocation points increased up to 512, and we found that a grid of 96 points in the $z$-direction is sufficient to accurately obtain the maximal growth rate. We then use a grid of 128 to implement our entire linear stability analysis in the range of $Ra\in[1, 10^7]$ and $Ri\in[0.001,10]$ for fixed parameters $(Pr, \tau, \Lambda)=(7, 0.01, 2)$. 100 uniform sample points are used for both $Ra$ and $Ri$. To find the maximal growth rate for each parameter set, we use a range of horizontal wavenumbers $k_x, k_y\in[10^{-3},1]$ with 100 uniform sample points for each direction. Our results indicate that the base state is linearly stable within this parameter regime.

\section{Linear stability analysis of vertically periodic SDDC at the $Pr\rightarrow \infty$ limit}\label{app:LAS_stokes_model}

In a vertically periodic domain with a sinusoidal shear flow, SDDC has thermohaline-shear instability \citep{radko2016thermohaline}. In this section, we investigate thermohaline-shear instability in vertically periodic SDDC at the limit of infinite Prandtl number, which relates to the $Pr$-effect (figure \ref{fig:tau_continuation}(b)) discussed in section \ref{sec:parameter_continuation}. By following the work of \citet[equation 4.1]{radko2016thermohaline}, we consider a sinusoidal background velocity profile $\boldsymbol{U}_b=\sin(2 \pi z)\hat{\mathbf{e}}_x$ and a negative gradient of temperature and salinity such that $T_b=-z$ and $S_b=-\Lambda z$, leading to the linearized non-dimensional governing equations of vertically periodic DDC in the diffusive regime \citep{radko2016thermohaline} as
\begin{subequations}
\label{eq:vertically_periodic_ddc}
\begin{align}
    \nabla \cdot \boldsymbol{u} &= 0,\\
    \frac{\partial u}{\partial t} &= -U_b\frac{\partial u}{\partial x} - w\frac{d U_b}{d z} -\frac{\partial p}{\partial x} + \frac{Pr}{Pe} \nabla^2 u, \label{eq:radko_2016_momentum.1}\\
    \frac{\partial v}{\partial t} &= - U_b \frac{\partial v}{\partial x} -\frac{\partial p}{\partial y} + \frac{Pr}{Pe} \nabla^2 v,\label{eq:radko_2016_momentum.2}\\
    \frac{\partial w}{\partial t} &= - U_b \frac{\partial w}{\partial x} -\frac{\partial p}{\partial z} + \frac{4 \pi^2 Ri}{\Lambda - 1} (\theta - s) + \frac{Pr}{Pe} \nabla^2 w, \label{eq:radko_2016_momentum.3}\\
    \frac{\partial \theta}{\partial t} &= - U_b \frac{\partial \theta}{\partial x} + w + \frac{1}{Pe} \nabla^2 \theta, \label{eq:radko_2016_temperature}\\
    \frac{\partial s}{\partial t} &= - U_b \frac{\partial s}{\partial x} + \Lambda w + \frac{\tau}{Pe} \nabla^2 s, \label{eq:radko_2016_salinity}
\end{align}
\end{subequations}
with $U_b=\sin(2 \pi z)$. Note that these governing equations \eqref{eq:vertically_periodic_ddc} are non-dimensionalized by the amplitude $U_0$ of the dimensional sinusoidal velocity profile $\boldsymbol{U}_b^*=U_0\sin(2 \pi z^*/H)\hat{\mathbf{e}}_x$, while equations \eqref{eq.sddc_fluctuation} are non-dimensionalized by the free-fall velocity. Also, both temperature and salinity are normalized by the temperature properties, leading to the presence of the density ratio $\Lambda$ in equation \eqref{eq:radko_2016_salinity}. The Peclet number ($Pe$) and the Richardson number are defined by 
\begin{equation}
    Pe=\frac{U_0 H}{\kappa_T}, \quad Ri=\frac{g\beta_T |\partial_{z^*}
 T^*| H^2}{4\pi^2 U_0^2}(\Lambda-1).
\end{equation}
The Rayleigh number $Ra$ can be determined through  $Ra=\frac{4\pi^2 Pe^2 Ri}{Pr(\Lambda-1)}$. The momentum equations \eqref{eq:radko_2016_momentum.1}-\eqref{eq:radko_2016_momentum.3} can be rewritten in the form of 
\begin{subequations}
\label{eq:vertically_periodic_ddc.2}
\begin{align}
    \frac{1}{Pr}\left(\frac{\partial u}{\partial t} +U_b\frac{\partial u}{\partial x} + w\frac{d U_b}{d z}\right) &=  -\frac{\partial p}{\partial x} + \frac{1}{Pe} \nabla^2 u, \label{eq:rewritten_radko_2016_momentum.1}\\
    \frac{1}{Pr}\left(\frac{\partial v}{\partial t} + U_b \frac{\partial v}{\partial x} \right)&= -\frac{\partial p}{\partial y} + \frac{1}{Pe} \nabla^2 v,\label{eq:rewritten_radko_2016_momentum.2}\\
    \frac{1}{Pr}\left(\frac{\partial w}{\partial t} + U_b \frac{\partial w}{\partial x} \right)&= -\frac{\partial p}{\partial z} + \frac{Ra}{Pe^2} (\theta - s) + \frac{1}{Pe} \nabla^2 w\label{eq:rewritten_radko_2016_momentum.3}
\end{align}
\end{subequations}
after a suitable rescaling of pressure $p$. Then we consider the limit of $Pr\rightarrow \infty$, while $Pe$ and $Ra$ are fixed. This limit will remove the inertial term in the momentum equation, i.e., the left-hand side of \eqref{eq:vertically_periodic_ddc.2} becomes zero. The numerical setups and the establishment of the matrix form are similar to those in Appendix \ref{app:LSA_base_SDDC}, except that we use the Fourier differentiation matrix in the vertical direction \citep{weideman2000matlab} due to the vertically periodic domain setup here. The eigenvalue problem is then solved to form the maximal growth rate map of ($Ri$, $Pe$). To find the maximal growth rate for each parameter set, we use a range of streamwise wavenumbers $k_x\in[10^{-3},2]$ with 100 uniform sample points. Since the 2D configuration exhibits a larger growth rate than the 3D case \citep{radko2016thermohaline}, we set the spanwise wavenumber to $k_y = 0$ to reduce computational cost. Subsequently, a map of the maximal growth rate is built with ranges of $Ri\in[0.25,10^3]$ and $Pe\in[1,10^4]$, comprising 100 logarithmically sampled points for each parameter.

\begin{figure}
    \centering
    \includegraphics[width=0.8\linewidth]{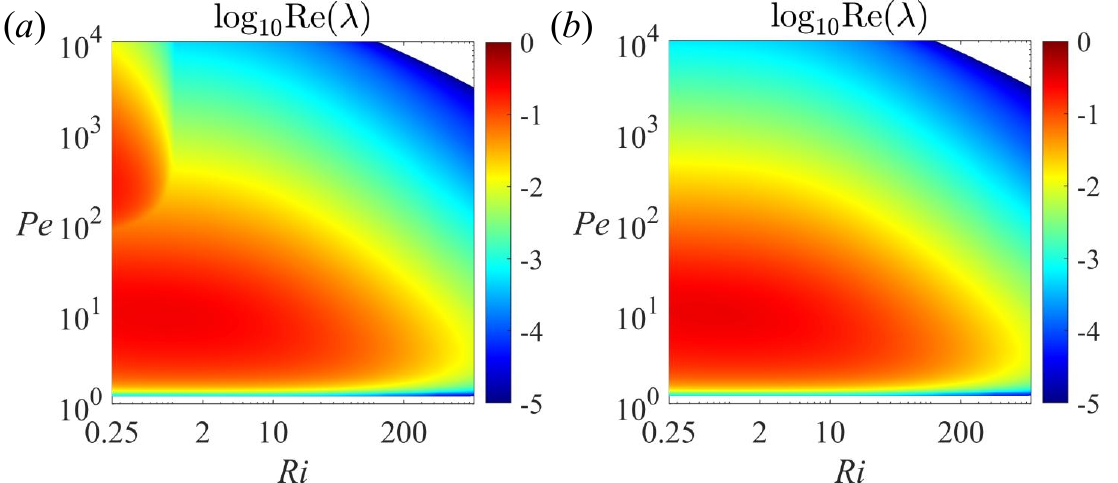}
    \caption{Maximal growth rate of linear stability analysis: (a) the full DDC model by \eqref{eq:vertically_periodic_ddc} \citep[equation 4.1]{radko2016thermohaline} at fixed parameters $(Pr, \Lambda, \tau)=(10, 2, 0.01)$ and (b) the reduced model without the inertial terms in the left-hand side of \eqref{eq:vertically_periodic_ddc.2} corresponding to the infinite Prandtl number limit. The result is plotted in the parameter space of $(Pe,Ri)$ with only growth rates exceeding $10^{-5}$.}
    \label{fig:ddc_radko2016}
\end{figure}

Figure \ref{fig:ddc_radko2016} shows a comparison of the maximal growth rate of linear stability analysis for the full SDDC model by \citet[equation 4.1]{radko2016thermohaline} and the model without the inertial term in the momentum equation at $Pr\rightarrow \infty$ limit (i.e., the left-hand side of \eqref{eq:vertically_periodic_ddc.2} is zero). Figure \ref{fig:ddc_radko2016}(a) shows the maximal growth rate over ($Ri$,$Pe$) for the full SDDC model in a vertically periodic domain, reproducing results reported by \citet[Figure 10b]{radko2016thermohaline}. Here, parameters $(Pr,\tau,\Lambda)=(10,0.01,2)$ were used. The plot indicates that the unstable properties may be split into two major instabilities, including thermohaline-shear instability on almost the whole map and shear instability in a small area where $Pe>10^2$ and $Ri<1$. To explore the origins of these instabilities, figure \ref{fig:ddc_radko2016}(b) shows the maximal growth rate of the reduced SDDC model by considering the $Pr\rightarrow \infty $ limit, removing the inertial term in the momentum equation, which filters out the possibility of shear instability. Other parameters ($\Lambda, \tau, Pe, Ra$) are the same as those in figure \ref{fig:ddc_radko2016}(a). As shown in figure \ref{fig:ddc_radko2016}(b), the shear instability associated with the $Ri<1$ regime disappears as expected because the inertial terms in the momentum equation are zero at this limit. This demonstrates that the thermohaline-shear instability arises from the interaction of thermohaline convection with the background shear as discussed by \citet{radko2016thermohaline}, which continues to exist even when we filter out the inertial term associated with the $Pr\rightarrow \infty$ limit. This differs from the $Pr$-effect (figure \ref{fig:tau_continuation}b) of SDDC within a vertically bounded domain with a linear background flow discussed in section \ref{sec:parameter_continuation}, where the thermohaline-shear instability does not occur at the high Prandtl number limit.

\bibliographystyle{jfm}
\bibliography{refs}
\end{document}